# Exact solution for the Anisotropic Ornstein-Uhlenbeck Process

Rita M.C. de Almeida[1,2,3*], Guilherme S. Y. Giardini[1], Mendeli Vainstein[1], James A. Glazier[4], and Gilberto L. Thomas[1*]

[1]*Instituto de Física* and [2]*Instituto Nacional de Ciência e Tecnologia: Sistemas Complexos,*
*Universidade Federal do Rio Grande do Sul, Porto Alegre, RS, Brazil*

[3]*Program de Pós Graduação em Bioinformática,*
*Universidade Federal do Rio Grande do Norte, Natal, RN, Brazil*

[4]*Biocomplexity Institute and Department of Intelligent Systems Engineering, Indiana University, Bloomington, Indiana, United States of America*

*Corresponding authors

Email: rita@if.ufrgs.br, glt@if.ufrgs.br

Abstract

Active-Matter models commonly consider particles with overdamped dynamics subject to a force (speed) with constant modulus and random direction. Some models also include random noise in particle displacement (a Wiener process) resulting in diffusive motion at short time scales. On the other hand, Ornstein-Uhlenbeck processes apply Langevin dynamics to the particles' velocity and predict motion that is not diffusive at short time scales. Experiments show that migrating cells have gradually varying speeds at intermediate and long time scales, with short-time diffusive behavior. While Ornstein-Uhlenbeck processes can describe the moderate- and long-time speed variation, Active-Matter models for over-damped particles can explain the short-time diffusive behavior. Isotropic models cannot explain both regimes, because short-time diffusion renders instantaneous velocity ill-defined, and prevents the use of dynamical equations that require velocity time-derivatives. On the other hand, both models correctly describe some of the different temporal regimes seen in migrating biological cells and must, in the appropriate limit, yield the same observable predictions. Here we propose and solve analytically an Anisotropic Ornstein-Uhlenbeck process for polarized particles, with Langevin dynamics governing the particle´s movement in the polarization direction and a Wiener process governing displacement in the orthogonal direction. Our characterization provides a theoretically robust way to compare movement in dimensionless simulations to movement in experiments in which measurements have meaningful space and time units. We also propose how to deal with inevitable finite precision effects in experiments and simulations.

# 1 Introduction

Observation and quantification of single-cell migration on flat surfaces dates back over a century [1,2]. Such cell movement has been often described by a Fürth equation that gives a cell´s Mean-Squared Displacement ($MSD$) as a function of the time interval $\Delta t$ between the acquisition of the cell´s positions used to calculate displacement:

$$MSD_{\text{Fürth}} = 4D\big[\Delta t - P\big(1 - \exp(-\Delta t/P)\big)\big] \,, \qquad (1)$$

where $D$ is the diffusion coefficient (for long time intervals, $MSD_{\text{Fürth}} \sim 4D\Delta t$) with the factor 4 accounting for movement in two dimensions. Over short time intervals, $MSD_{\text{Fürth}} \sim \frac{2D}{P}\Delta t^2$ and motion is ballistic, allowing consistent definition of the instantaneous velocity (which is ill-defined for a Wiener process). The persistence time, $P$, is the time interval at which the movement transitions from ballistic to diffusive [3,4,5,6,7,8]. Eq. (1) is the solution of an Ornstein-Uhlenbeck process for particle motion; that is,

$$\frac{d\vec{v}}{dt} = -\gamma\vec{v} + \vec{\xi}(t) \qquad (2a)$$

$$\frac{d\vec{r}}{dt} = \vec{v} \,, \qquad (2b)$$

where $\vec{r}$ and $\vec{v}$ are, respectively, the particle´s position and instantaneous velocity, $\gamma$ is the strength of dissipation, that consumes kinetic energy, and $\vec{\xi}(t)$ is a two-dimensional white noise vector from which the particle gathers kinetic energy. Trajectories obtained from solving Eqs. (2), allow calculation of the $MSD$. Classical Brownian particles at a liquid surface obey the same set of equations, where $\gamma$ is the fluid viscosity and $\vec{\xi}(t)$ describes the impulse the particle receives from collisions with fluid molecules. Since migrating cells are neither isotropic nor inert particles set in movement by interaction with the thermal motion of the components of their environment, describing cell movement requires the reinterpretation of each term in Eqs. (2). Cell trajectory $MSD$s deviate from the Fürth equation, requiring additional adjustments. Thomas and collaborators [9] demonstrated that eukaryotic single-cell migration shows Ornstein-Uhlenbeck-like statistics for intermediate- and long-time scales but diffusive statistics for short-time scales. Because the instantaneous velocity of the cells is divergent, the inferred velocity and diffusion constant depend on the time interval between position measurements, impeding consistent comparisons between experiments. Computer simulations of 3D crawling cells using the Cellular Potts Model in CompuCell3D also show short-time diffusive movement [10]. Since experiments and simulations necessarily have some shortest interval between position measurements, we need metrics to quantify movement that are independent of this minimum. Another valuable tool to investigate cell migration is the Velocity Auto-Correlation Function (*VACF*), defined as the average scalar product of the velocity at a given time with the velocity after a time interval $\Delta t$. For stationary processes, the *VACF* is the second time derivative of the *MSD*. However, when time intervals are small, the inevitable finite precision in measurements leads to a marked decrease in the modulus of the *VACF*, compared to the values predicted by the *MSD*'s second time derivative. Such velocity correlation loss will occur in any system that with similar short-time diffusive behavior [9].

Active-Matter models have also been applied to model migrating biological cells [11]. In some of these models, the particle´s speed $v_0$ is constant, while its direction may change due to a white noise term [12]. In these cases, the cell movement is modelled by overdamped dynamics at small Reynolds numbers, where drag instantaneously eliminates speed deviations from the constant value defined by the balance between drag and the internal mechanisms responsible for cell movement [11,12,13,14]. In these models *MSD* obeys Eq. (1). The biological interpretation is that the particle speed is due to internal activity of the cell. This internal activity ceases at random times, the speed goes to zero over a negligible (infinitesimal) time interval and the internal activity resumes instantaneously with the same modulus, but a different direction over the next time interval: the particle is active, but its dynamics is overdamped. The movement direction, denoted by an angle $\theta$ with respect to the reference frame, remembers the direction of movement during the previous time interval, changing stochastically by small amounts. An additional, white noise term ($\vec{\omega}(t)$) may also be added to the displacement equation; yielding,

$$v(t) = v_0 \quad (3a)$$
$$\frac{d\theta}{dt} = \beta(t) \quad (3b)$$
$$\frac{d\vec{r}}{dt} = v_0\vec{p}(t) + \vec{\omega}(t), \quad (3c)$$

where $\vec{r}$ is the particle's position, $\vec{p}(t) = (\cos\theta, \sin\theta)$ is the particle's polarization, and $\beta(t)$ and $\vec{\omega}(t)$ are white noise terms with appropriate units. When $\vec{\omega}(t)$ differs from zero, instantaneous velocity is not well-defined, since $\lim_{\Delta t \to 0} \frac{\vec{r}(t+\Delta t)-\vec{r}(t)}{\Delta t}$ diverges. In other words, $v_0$ is not given by the ratio of displacement to time interval in the limit of vanishing time intervals and is thus not a proper velocity, but rather a model parameter, associated with the cell's internal force-generation. Non-zero $\vec{\omega}(t)$ results in short-time-interval diffusion that translates into a $MSD \sim \Delta t$ as $\Delta t \to 0$.

Since Eqs. (3) do not include a velocity derivative, a non-zero $\vec{\omega}(t)$ leaves Eqs. (3) well behaved, although characterizing $v(t)$ requires new measurement protocols. Both the Ornstein-Uhlenbeck process (Eqs. (2)) and Active-Matter models for overdamped particles (Eqs. (3)) correctly describe some of the different temporal regimes seen in migrating biological cells [17,18,11] and must, in the appropriate limit, yield the same observable predictions. Neglecting both $\vec{\omega}(t)$ and the short-time diffusive regime in the *MSD* curves, we can relate the two sets of equations by changing the first equation in Eqs. (3) into the Orstein-Uhlenbeck equation for velocity (Eqs. (2)) [11,12]. On the other hand, if we want to account for the universally-observed diffusive regime for short-time intervals, $\vec{\omega}(t)$ is non-zero and the ill-defined instantaneous velocity prevents us from writing an equation that involves the velocity time-derivative as in Eqs. (2).

Another way to produce short-time-interval diffusive behavior MSD curves, is to follow Peruani and Morelli [11] and consider models with decoupled speed and orientation dynamics. The resulting *MSD* is a sum of two Fürth equations with different time-scales. In this case, the model predicts that for increasing time intervals, the motion transitions from ballistic to diffusive to ballistic and finally to diffusive regimes. For experiments with short (but not too short) time intervals between measurements, Peruani and Morelli´s model could describe observed diffusive deviations from the original Fürth equation. Motion in Peruani and Morelli´s model is isotropic, so all directional components of the velocity have equivalent short-time-interval diffusive motion.

Here we approach cell motion by explicitly considering the experimentally-observed anisotropy of migrating cells [19]. As we explain in the next section, we assign to the particle a polarization degree of freedom. The polarization direction continuously changes as described by the $\theta$-equation in Eqs. (3), and, at each instant, the particle's speed in the polarization direction obeys an Ornstein-Uhlenbeck process, while in the orthogonal direction(s) the particle's displacement obeys a Wiener process. Below, we propose and analytically solve this mixed model. We cannot use Peruani and Morelli´s formalism to obtain *MSD* curves, since our model couples speed and orientation dynamics, so we have developed a different approach. We show that the *MSD* curves in this model have a short-time-scale diffusive regime as do Active-Matter models with non-zero $\vec{\omega}(t)$ [11,12,13] and eukaryotic migrating cells [7,9]. We also show that the translational-noise anisotropy affects the definition of cell speed and the protocol needed to measure it. Furthermore, we predict that the fast-diffusive regime is present only for movement in the direction orthogonal to the polarization direction, which allows measurements to discriminate between our anisotropic model and Peruani-like dynamics as a mechanistic explanation for the observed fast-diffusive dynamics. We also numerically solve the model equations, to verify the analytical solutions and obtain representative trajectories. Finally, we show how finite precision in numerical solutions or in experimental measurements can lead to deviations from the theoretical predictions for the *VACF* for short-time intervals.

## 2 The Anisotropic Ornstein-Uhlenbeck model

We assume that a particle has an internal orientational degree of freedom, given by a polarization vector, $\vec{p}(t) = (cos\,\theta(t), sin\,\theta(t))$. In a biological cell, this orientation might define the direction of cell polarization or planar polarity [14], in an animal, the vector pointing from tail to head. We alternate changes in the direction of polarization with changes in speed in a fixed direction of polarization. We begin by defining polarization dynamics as:

$$[\theta(t + \Delta t) - \theta(t)] = \int_t^{t+\Delta t} \beta_\perp(t)\,dt, \qquad (4)$$

where $\beta_\perp(t)$ is a Gaussian white noise. The statistics of movement parallel and perpendicular to $\vec{p}(t)$ differ. In the polarization direction, we assume that for a small-time interval $\Delta t$ we may write the change in the magnitude of the cell´s velocity (which we will call the *parallel velocity*):

$$v_\parallel^{final}(t) = \left[(1 - \gamma\Delta t)v_\parallel^{initial}(t) + \int_t^{t+\Delta t} \xi_\parallel(t)dt\right], \qquad (5)$$

where $\gamma$ is the dissipation and $\xi_\parallel(t)$ is also a Gaussian white noise, with appropriate units. $v_\parallel^{initial}(t)$ and $v_\parallel^{final}(t)$ are the magnitudes of the parallel velocities, respectively, at the beginning and at the end of the time interval $\Delta t$. At the end of that small time interval, we assume that the polarization direction changes, from $\vec{p}(t)$ to $\vec{p}(t + \Delta t)$, and the initial parallel velocity at the beginning of the subsequent time interval is the projection of $v_\parallel^{final}(t)\vec{p}(t)$ onto $\vec{p}(t + \Delta t)$, that is:

$$v_\parallel^{initial}(t + \Delta t) = v_\parallel^{final}(t)\big(\vec{p}(t) \cdot \vec{p}(t + \Delta t)\big)\,. \qquad (6)$$

In Eq. (6) we hypothesize that the actin-filament dynamics is subject to noise that may randomly reorient the rear-to-front axis that defines a migrating cell's polarization, obtained from Eq. (4). We also hypothesize that these direction changes reduce cell speed, since a migrating cell´s speed is universally coupled to its cytoskeletal organization [21]. Here we assume that we may describe the conserved fraction of speed (the speed 'memory') by the projection of the new polarization direction onto the previous one. Eqs. (5) and (6) assume Itô integration of the stochastic variables, without anticipation. Eq. (6) couples the dynamics of the migration orientation $\theta(t)$ to those for the parallel velocity $v_\parallel(t)$: thus Peruani´s assumptions do not hold [11]. Eqs. (5) and (6) yield an evolution equation for the parallel velocity, $v_\parallel(t)$, which is well-defined:

$$v_\parallel(t+\Delta t)\,\vec{p}(t+\Delta t) = \left[(1-\gamma\Delta t)v_\parallel(t) + \int_t^{t+\Delta t}\xi_\parallel(t)dt\right](\vec{p}(t)\cdot\vec{p}(t+\Delta t))\,\vec{p}(t+\Delta t), \quad (7)$$

Taking the limit as $\Delta t \to 0$ with alternating steps for orientation and parallel-velocity changes might seem problematic. However, while the dynamics of the polarization direction, $\vec{p}(t) = (\cos\theta(t), \sin\theta(t))$, follow Eq. (4), a Wiener process for which the variables are not constant even over an infinitesimal time interval, Eq. (7) considers only $\vec{p}(t)\cdot\vec{p}(t+\Delta t) = \cos\Delta\theta(t) \sim 1 - \frac{(\Delta\theta)^2}{2}$, so $\langle(\Delta\theta)^2\rangle \sim \Delta t$. Hence, in Eq. (7), we can assume that $\vec{p}(t)$ is constant in the limit that $\Delta t$ is small. In the supplementary materials online, we justify in detail our assumption of an infinitesimal time interval for $v_\parallel(t)$ dynamics in Eq. (7).

The particle **position** in the direction orthogonal to the polarization obeys a Wiener process:

$$[r_\perp(t+\Delta t) - r_\perp(t)]\,\vec{n}(t) = \vec{n}(t)\int_t^{t+\Delta t}\xi_\perp(t)\,dt, \quad (8)$$

where $\vec{n}(t) = (\sin(\theta(t)), -\cos(\theta(t)))$ is a unit vector perpendicular to $\vec{p}(t)$. This change happens during the time interval as in Eq. (7), between the rotations described by Eq. (6).

$\xi_\parallel(t)$, $\xi_\perp(t)$, and $\beta_\perp(t)$ are all Gaussian white noise (with different units, see below). $\xi_\parallel(t)$ is independent of the two other terms, but we assume that $\xi_\perp(t)$ and $\beta_\perp(t)$ are related because fluctuations in the actin-network dynamics in the lamellipodium are responsible for both stochastic change in the rear-to-front direction, and for random displacements in the $\vec{n}(t)$ direction. We assume:

$$\xi_\perp(t) = \sqrt{q}\beta_\perp(t), \quad (9)$$

with $\sqrt{q}$ given in units of length. The noise terms are given in terms of their second moments:

$$\langle\xi_\parallel(t)\rangle = 0, \qquad \langle\xi_\parallel(t)\xi_\parallel(t')\rangle = g\,\delta(t-t'), \quad (10a)$$
$$\langle\beta_\perp(t)\rangle = 0, \qquad \langle\beta_\perp(t)\beta_\perp(t')\rangle = 2k\,\delta(t-t'), \quad (10b)$$
$$\langle\xi_\perp(t)\rangle = 0, \qquad \langle\xi_\perp(t)\xi_\perp(t')\rangle = 2qk\,\delta(t-t'), \quad (10c)$$

where $g$, $k$, and $qk$ have units of $[length]^2/time^3$, $1/time$ and $[length]^2/time$, respectively.

We summarize our model in Fig.1: it considers a particle with two spatial degrees of freedom and one internal polarization degree of freedom that breaks the cell´s spatial symmetry. The particle follows a Langevin-like dynamics for speed in the instantaneous polarization direction and, in the direction perpendicular to the instantaneous polarization, a Wiener process for displacement. The polarization direction also follows a Wiener process. Of the two independent sources of noise, one changes the speed in the polarization direction and the second both changes the polarization direction and applies a random displacement in the direction orthogonal to the polarization. The change in polarization causes loss of time correlation in the velocity and, as we show below, reduces the persistence time of the movement.

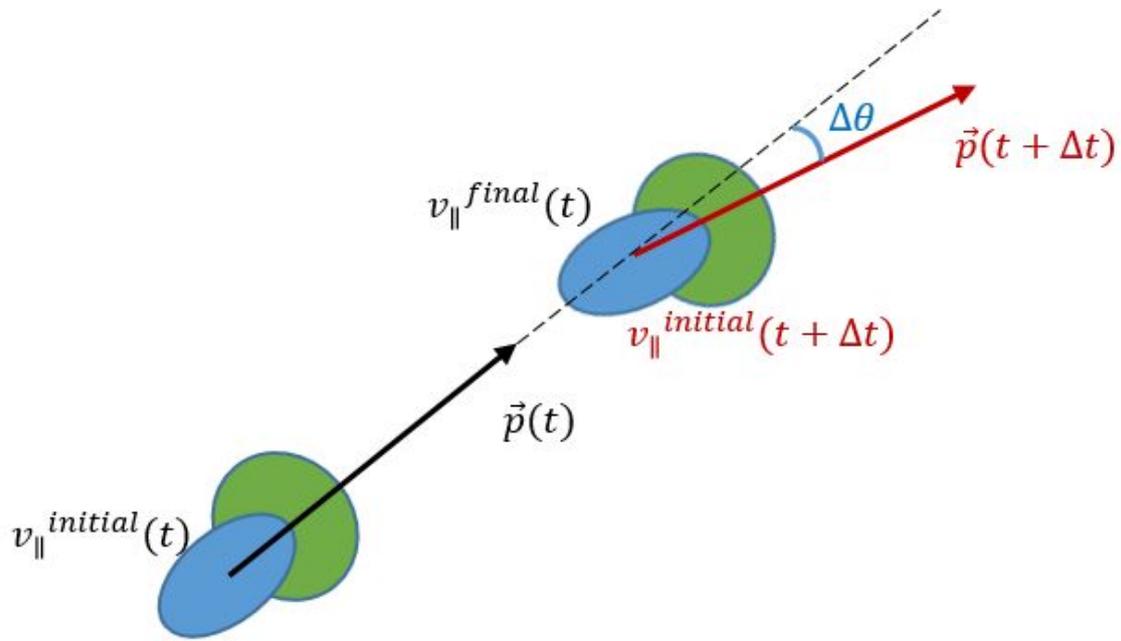

*Figure 1. Sketch of the model. At the beginning of a small-time interval, $\Delta t$, the cell has initial parallel velocity $v_\parallel^{initial}(t)$ and polarization direction $\vec{p}(t)$. At the end of the interval the parallel velocity changes to $v_\parallel^{final}(t)$ following Eq. (5), after which the polarization changes to $\vec{p}(t + \Delta t)$, with $\vec{p}(t) \cdot \vec{p}(t + \Delta t) = \cos \Delta\theta$, with $\Delta\theta$ evolving according to Eq. (4) (a Wiener process). At the beginning of the next step, the change of polarization reduces the parallel velocity from $v_\parallel^{final}(t)$ to $v_\parallel^{initial}(t + \Delta t)$ (Eq. (6)). We also assume a random displacement in the direction perpendicular to the polarization axis during $\Delta t$, before the change in polarization direction (Eq. (8)).*

**3 Numerical solutions for the Anisotropic Ornstein-Uhlenbeck Process**

To explore our analytic results, we solved the dynamics represented by Eqs. (3)-(6) numerically. These equations initially have 4 parameters: $q$, $\gamma$, $g$, and $k$. We wrote a C language program using the Euler-Maruyama method for integrating stochastic differential equations [15]. When we analyze the movement (below) we find that by rescaling the parallel and perpendicular length scales and the time scale we can eliminate three parameters, leaving the single parameter $k$. As we show analytically below, solving Eqs. (4), (7), and (8) yields *MSD* curves that reproduce the Modified Fürth Equation (Eq. (22) below), empirically-proposed in Ref. [9]. The Modified Fürth Equation, when written using non-dimensional variables, represents a single-parameter family of curves, where the single parameter $S$ (the *excess diffusion coefficient*) defines the time duration of the short-time-diffusive regime. $MSD_{\text{Fürth}}$ given in Eq. (1), is the member of this family of curves with $S = 0$. Eqs. (23) relates $q$, $\gamma$, $g$, and $k$ to the observable parameters $S$, $P$, and $D$, while the length and time scales are $\sqrt{\frac{2DP}{1-S}}$ and $P$, as in Ref. [9].

Without loss of generality, we consider all parameters except for $k$ constant ($q = 0.1$, $\gamma = 1$ and $g = 10$): varying $k$ over the range $[0.04, 2.0]$ ($k \in \{0.04405, 0.2625, 0.965, 1.7425\}$, which correspond to $S \in \{0.001, 0.01, 0.1, 0.3\}$), exhibits all experimentally-observed dynamics of migrating cells. The other parameters define the length and time scales of the measurements.

As in an ordinary Langevin problem, our model admits a stationary state, in which the average speed, *MSD* and *VACF* curves do not change in time. Initial cell polarization angles are randomly and uniformly distributed in $[0, 2\pi)$ and $v_\parallel(t = 0)$ is initialized either to the parallel velocity in the stationary state $v_\parallel(t = 0) = \sqrt{\langle v_{\parallel sta}^2 \rangle}$ (Eq. (16), below), to show the stationary *MSD* and *VACF*, or $v_\parallel(t = 0) = 10^3$, to show how the transient relaxes to the stationary state.

Each time step of the dynamics consists of the following substeps: *i)* we choose a Gaussian random number with standard deviation equal to $g\,\Delta t$ and update $v_\parallel$ according to Eq. (5); *ii)* we choose an independent Gaussian random number with standard deviation equal to $2kq\,\Delta t$ and determine the perpendicular displacement (Eq. 8); *iii)* we update the cell position; iv) we update the polarization angle $\theta$ according to Eq. (4) and (9); *v)* we project $v_\parallel$ onto the new direction, according to Eq. (6). We repeat these steps $10^6$ times (we used $\Delta t = 10^{-4}$) and we average over 100 independent cells.

Figure 2. Trajectories for the parameter set $q = 0.1, g = 10, \gamma = 1$, with $k$ as indicated in the panels, where the particle´s position $\vec{\rho} = (\rho_x, \rho_y)$ is given in terms of the natural length unit

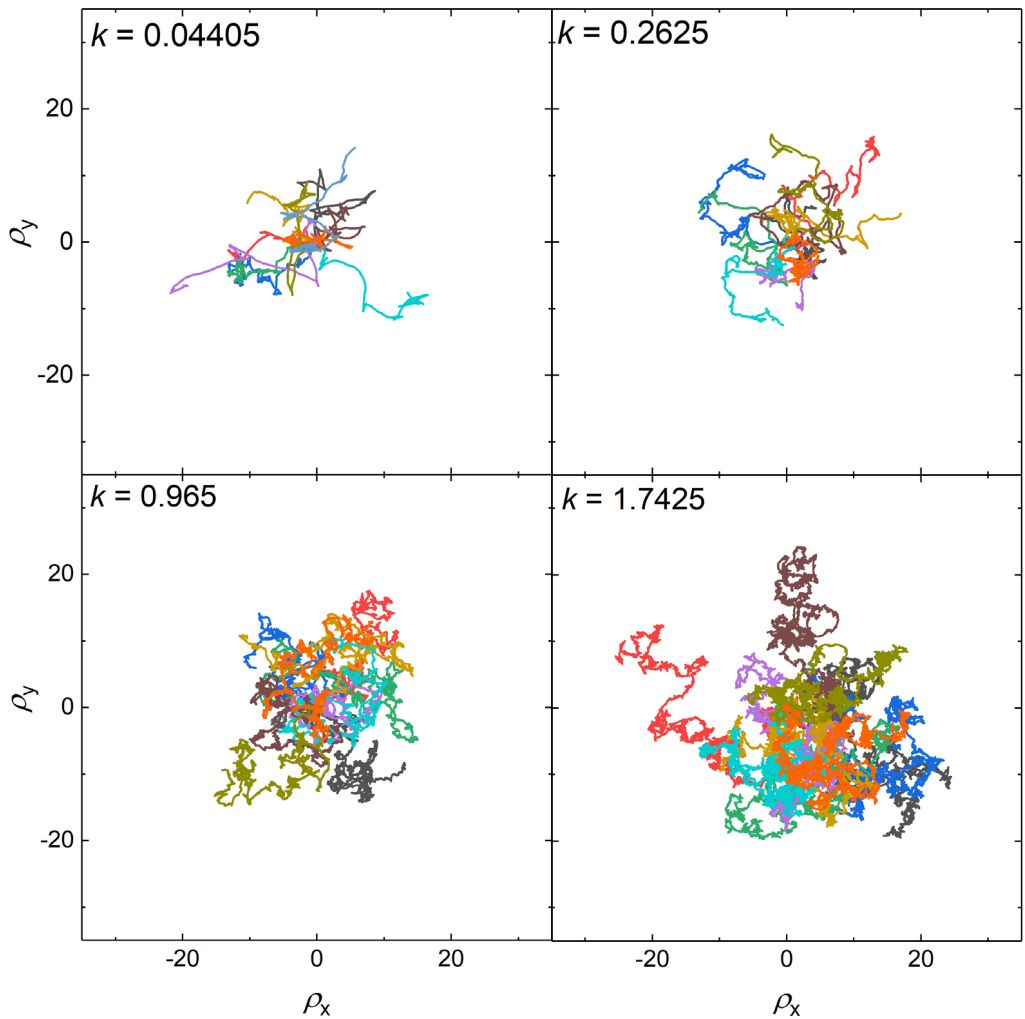

$\left(\sqrt{\frac{2DP}{1-S}}\right)$. Each panel shows 10 trajectories of $10^6$ steps, randomly chosen from the 100 trajectories calculated for each parameter set. Larger values of $k$ yield more convoluted trajectories (shorter persistence lengths).

## 4 Analytical solutions for *MSD* and *VACF*.

Below, we present exact solutions for this model's *MSD* and *VACF*. We obtained our analytical solutions at time $T$ by considering $n$ steps, each of duration $\Delta t = \frac{T}{n}$, then taking the limit $\Delta t \to 0$, while $n = \frac{T}{\Delta t} \to \infty$, so $T$ remains constant.

### 4.1 Analytical forms for $\langle v_\parallel^2(n\Delta t)\rangle$ and the persistence time *P.*

In what follows, we define $\vec{p}_j \equiv \vec{p}(j\Delta t)$. We apply Eq. (7), to obtain the parallel velocity. We first calculate $v_\parallel(\Delta t)\, \vec{p}(\Delta t)$ in terms of $v_\parallel(0)\, \vec{p}(0)$:

$$v_\parallel(\Delta t)\vec{p}(\Delta t) = \left[(1-\gamma\Delta t)v_{\parallel_0} + \int_0^{\Delta t}\xi_\parallel(t)\,\mathrm{d}t\right](\vec{p}_0\cdot\vec{p}_1)\,\vec{p}_1. \tag{11}$$

We then iterate Eq. (7) $n = \frac{T}{\Delta t}$ times to obtain $v_\parallel(T)\vec{p}(T)$ in terms of $v_\parallel(0)\,\vec{p}(0)$:

$$v_\parallel(n\,\Delta t)\vec{p}_n = (1-\gamma\Delta t)^n\,v_{\parallel_0}\prod_{i=0}^{n-1}[\vec{p}_i\cdot\vec{p}_{i+1}]\,\vec{p}_n$$

$$+\sum_{j=0}^{n-1}\int_{j\Delta t}^{(j+1)\Delta t}\mathrm{d}s\,\xi_\parallel(s)\,(1-\gamma\Delta t)^{n-(j+1)}\prod_{i=j}^{n-1}[\vec{p}_i\cdot\vec{p}_{i+1}]\,\vec{p}_n. \tag{12}$$

From Eq. (12), we calculate $\langle v_\parallel^2(n\Delta t)\rangle$ as follows:

$$\langle v_\parallel^2(n\Delta t)\rangle = (1-\gamma\Delta t)^{2n}v_{\parallel_0}^2\left\langle\prod_{i=0}^{n-1}[\vec{p}_i\cdot\vec{p}_{i+1}]^2\right\rangle$$

$$+g\sum_{j=0}^{n-1}\int_{j\Delta t}^{(j+1)\Delta t}\mathrm{d}s\,(1-\gamma\Delta t)^{2(n-(j+1))}\left\langle\prod_{i=j}^{n-1}[\vec{p}_i\cdot\vec{p}_{i+1}]^2\right\rangle, \tag{13}$$

where we used Eq. (10a) for the average over $\xi_\parallel(t)$. To calculate the average over the stochastic changes in $\vec{p}_i$, we note that $[\vec{p}_{j-1}\cdot\vec{p}_j] = \cos\bigl(\theta((j-1)\Delta t) - \theta(j\Delta t)\bigr) = \cos(\Delta\theta)$. For small $\Delta t$, $\cos(\Delta\theta) \sim 1 - \frac{1}{2}(\Delta\theta)^2$ and $\cos^2(\Delta\theta) \sim 1 - (\Delta\theta)^2$. Using Eq. (10b), we have $\langle(\Delta\theta)^2\rangle = 2k\Delta t$ and

$$\langle v_\parallel^2(n\Delta t)\rangle = v_{\parallel_0}^2\,(1-\gamma\Delta t)^{2n}(1-k\Delta t)^{2(n-1)}$$

$$+g\bigl[(1-\gamma\Delta t)^{2(n-1)}(1-k\Delta t)^{2(n-1)} + \cdots + 1\bigr]. \tag{14}$$

Taking the limit $\Delta t \to 0$, with $n = \frac{T}{\Delta t}$, we find:

$$\langle v_\parallel^2(T)\rangle = \frac{g}{2(\gamma+k)} + \left(v_{\parallel_0}^2 - \frac{g}{2(\gamma+k)}\right)\exp[-2(\gamma+k)T]. \tag{15}$$

If we assume the initial condition for the parallel velocity is the asymptotic solution, $v_{\parallel_0}^2 = \frac{g}{2(\gamma+k)}$, we find:

$$\langle v_{\parallel sta}^2 \rangle = \frac{g}{2(\gamma + k)}. \tag{16}$$

The *relaxation time R*, defined as:

$$R = (\gamma + k)^{-1}, \tag{17}$$

determines the rate at which the average squared speed approaches its asymptotic value. To compare with numerical solutions, we estimate the squared speed from the mean velocity over finite time intervals $\varepsilon$, that is, $\langle v_\parallel^2(T) \rangle \approx \langle \frac{|\vec{r}(T+\varepsilon)-\vec{r}(T)|^2}{\varepsilon^2} \rangle$, which, for small $\varepsilon$ decomposes into:

$$\langle \frac{|\vec{r}(T+\varepsilon)-\vec{r}(T)|^2}{\varepsilon^2} \rangle = \langle v_\parallel^2(T) \rangle + \langle \frac{|r_\perp(T+\varepsilon)-r_\perp(T)|^2}{\varepsilon^2} \rangle. \tag{18}$$

Figure 3 shows $\langle \frac{|\vec{r}(T+\varepsilon)-\vec{r}(T)|^2}{\varepsilon^2} \rangle - \frac{g}{2(\gamma+k)}$ as a function of time, for initial conditions with $v_{\parallel_0} = 10^3$ and different values of $k$. The symbols correspond to averages over numerically-calculated trajectories for each time $T$. The solid line is the analytical prediction, given by subtracting Eq. (16) from Eq. (15). Notice that $\lim_{T\to\infty}\left[\langle \frac{|\vec{r}(T+\varepsilon)-\vec{r}(T)|^2}{\varepsilon^2} \rangle - \frac{g}{2(\gamma+k)}\right] = \frac{2kq}{\varepsilon}$, as predicted if $\langle |r_\perp(T+\varepsilon) - r_\perp(T)|^2 \rangle = 2kq\varepsilon$.

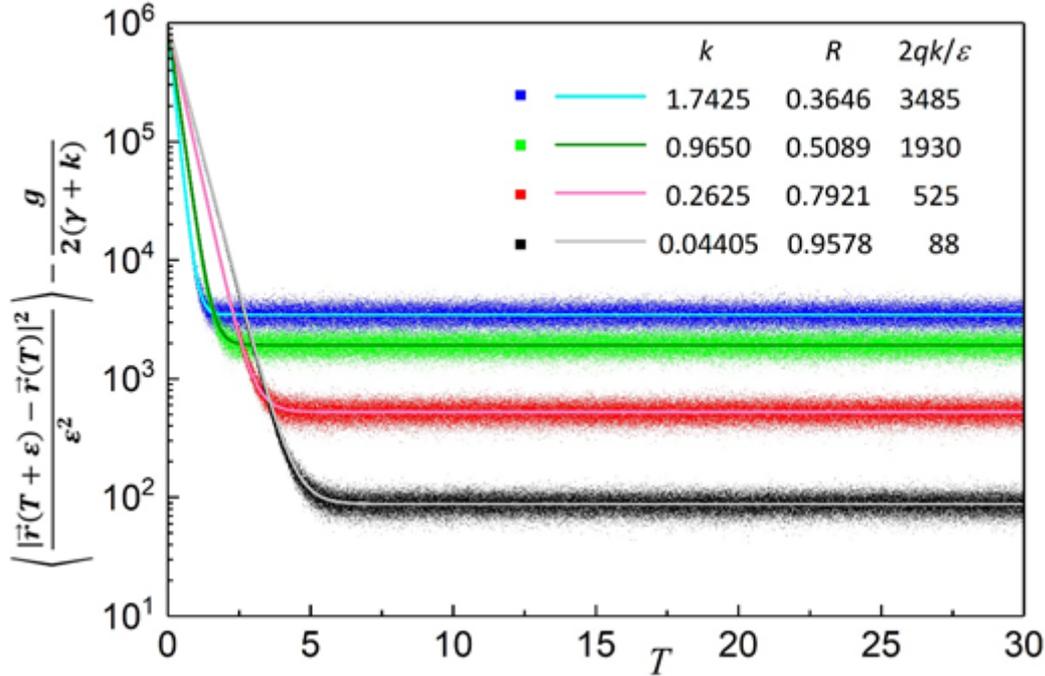

Figure 3. Semi-log plots of $\langle \frac{|\vec{r}(T+\varepsilon)-\vec{r}(T)|^2}{\varepsilon^2} \rangle - \frac{g}{2(\gamma+k)}$ versus T, for $\varepsilon = 10^{-4}$, q = 0.1, g = 10, $\gamma$ = 1, and k as indicated. R depends on k according to Eq. (17). Symbols correspond to estimates obtained from numerical iteration for 100 independent trajectories with $10^6$ iteration steps. Solid lines correspond to the analytical solutions obtained from Eq. (18).

Figure 4, shows the numerically-obtained probability density for the velocity parallel to the polarization, $F(u_{\parallel x}, u_{\parallel y})$, where $\vec{u}_\parallel = (u_{\parallel x}, u_{\parallel y})$ is the velocity parallel to the polarization given in terms of its components in the laboratory reference frame, measured in natural units of velocity, $\sqrt{2D/P(1-S)}$. D, P and S are functions of the model parameters $\gamma = 1, g = 10, k = 0.04405$, and $q = 0.10$ (see Eq.(23), below). The left panel shows $F(u_{\parallel x}, 0)$, while the right panel shows a heat map for the probability density function $F(u_{\parallel x}, u_{\parallel y})$. The right panel

shows that in the stationary state, the probability of finding the particle's polarization is the same for all orientations; the left panel shows that the probability diverges at the origin.

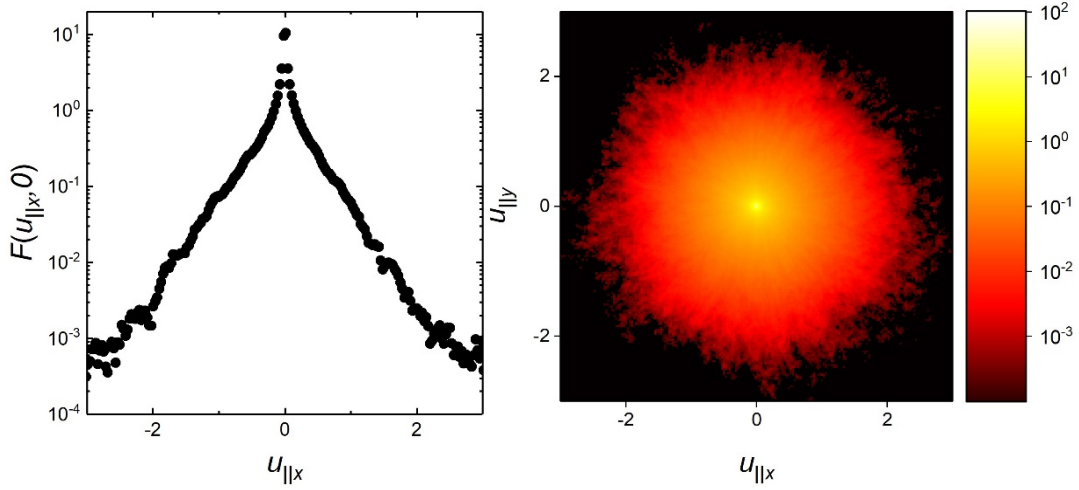

Figure 4. Probability density function for the velocity parallel to the polarization. Left Panel: histogram of $F(u_{\parallel x}, 0)$. Right Panel: heat map of the probability density function $F(u_{\parallel x}, u_{\parallel y})$ in the $(u_{\parallel x}, u_{\parallel y})$ plane. In both panels, $u_\parallel$ is $v_\parallel$ rescaled in natural units of velocity.

**4.2 Analytical forms for the Mean-Squared Displacement (MSD)**

We obtain the Mean-Squared Displacement by first calculating the displacement in each time interval $\Delta t$, from $T = 0$ to $T = n\,\Delta t$, then summing over the displacements, taking the square of this expression, and finally averaging over different trajectories, which is equivalent to the average over noise terms, since we consider the stationary solution. The Supplementary Materials Online provide details on these calculations.

After $n$ iterations ($n > 0$) the particle's displacement is:

$$\vec{r}(n\Delta t) - \vec{r}(0) = v_{\parallel_0}\Delta t \left[ \vec{p}_0 + \Theta(n-1) \sum_{j=0}^{n-1} (1-\gamma\Delta t)^{n-j-1} \prod_{m=0}^{n-j-2} [\vec{p}_m \cdot \vec{p}_{m+1}]\, \vec{p}_{n-j-1} \right]$$

$$+ \Delta t\, \Theta(n-2) \sum_{j=0}^{n-2} \int_{j\Delta t}^{(j+1)\Delta t} ds\xi_\parallel(s) \sum_{i=0}^{n-j-2} (1-\gamma\Delta t)^{n-i-j-2} \prod_{m=j}^{n-i-2} [\vec{p}_m \cdot \vec{p}_{m+1}]\, \vec{p}_{n-i-1}$$

$$+ \sum_{j=0}^{n-1} \int_{j\Delta t}^{(j+1)\Delta t} ds\xi_\parallel(s) [(j+1)\Delta t - s]\vec{p}_j + \sum_{j=0}^{n-1} \int_{j\Delta t}^{(j+1)\Delta t} ds\, \xi_\perp(s)\vec{n}_j\,, \quad (19)$$

where $\Theta(n-2) = 0$ if $n < 2$ and $\Theta(n-2) = 1$ otherwise. Squaring Eq. (19) and averaging over noise, we get:

$$MSD = \langle |\vec{r}(T+\Delta T) - \vec{r}(T)|^2 \rangle =$$

$$\frac{g}{(\gamma+2k)(\gamma+k)}\left[\Delta T - \frac{1}{\gamma+2k}\left(1 - e^{-(\gamma+2k)\Delta T}\right)\right] + 2qk\Delta T\,. \quad (20)$$

The Fürth equation is the MSD for the Langevin equation:

$$MSD_{\text{Fürth}} = 2D\big[\Delta T - P\big(1 - e^{-\Delta T/P}\big)\big]. \tag{21}$$

We can rewrite Eq. (20) as a *modified* Fürth equation:

$$MSD_{\text{ModifiedFürth}} = 2D\left[\frac{\Delta T}{(1-S)} - P\big(1 - e^{-\Delta T/P}\big)\right], \tag{22}$$

as proposed by Thomas *et al.* [9], where we identify:

$$D = \frac{g}{2(\gamma + 2k)(\gamma + k)}, \tag{23a}$$

$$P = \frac{1}{\gamma + 2k}, \tag{23b}$$

and

$$S = \frac{2qk(\gamma + 2k)(\gamma + k)}{g + 2qk(\gamma + 2k)(\gamma + k)}. \tag{23c}$$

Active matter models which add noise to the displacement yield *MSD* curves isomorphic to Eq. (22) [18]. Models with isotropic noise added to the displacement cannot use velocity derivatives in their dynamical equations.

Unlike the classical Ornstein-Uhlenbeck process, in our model, the persistence time $P$ (Eq. (23b)) is not the same as the relaxation time $R$ (as defined in Eq. (17)). $S$ and $D$ depend on both relaxation times, as given in Eqs. (23).

When $k = 0$, our model yields one-dimensional Fürth equations for both *MSD* and $\langle v_\parallel^2(T)\rangle$ relaxation, with $S = 0$, $P = \frac{1}{\gamma}$ and $D = \frac{g}{2\gamma^2}$. When $k > 0$, but $q = 0$, our model's *MSD* curve is the same as that of the Fürth equation, but the $\langle v_\parallel^2(T)\rangle$ relaxation time $R$ differs from that for the isotropic Ornstein-Uhlenbeck process. For $k > 0$ and $q > 0$, our model's *MSD* and $\langle v_\parallel^2\rangle$ relaxation times both differ from those of the isotropic Ornstein-Uhlenbeck process.

As observed in Ref. [9], for small $\Delta T$, Eq. (22) yields:

$$\lim_{\Delta T \to 0} MSD_{\text{ModifiedFürth}} \sim \frac{2SD}{1-S} \Delta T, \tag{24}$$

indicating that at short-time intervals, the particle's motion is diffusive with an effective diffusion constant $D_{fast} = \frac{SD}{1-S} = qk$. For long-time intervals, we find:

$$\lim_{\Delta T \to \infty} MSD_{\text{ModifiedFürth}} \sim \frac{2D}{1-S} \Delta T, \tag{25}$$

indicating a long-time diffusive behavior, with an effective diffusion constant $D_{slow} = \frac{D}{1-S}$. Together, these diffusion constants indicate the physical meaning of the parameter $S$: $S = \frac{D_{fast}}{D_{slow}}$. Following Ref. [9], we call $S$ the *excess diffusion coefficient*. The $MSD_{\text{ModifiedFürth}}$ in Eq. (22) has three regimes: a fast-diffusive regime for short-time intervals ($\Delta T < SP$), a ballistic-like, intermediate-time-interval regime ($SP < \Delta T < P$), and a slow-diffusive, long-time-interval regime ($\Delta T > P$). Fortuna and collaborators [10] found in their numerical simulations that $S = \frac{D_{fast}}{D+D_{fast}}$, while we show that this behavior is an exact consequence of the definition of $D_{fast}$ and Eqs. (23).

Below, following Ref. [9], we use $\sqrt{\frac{2DP}{1-S}}$ as a length scale and $P$ as a time scale to rewrite Eq. (23) as:

$$\langle |\Delta\vec{\rho}|^2 \rangle = \Delta\tau - (1-S)(1 - e^{-\Delta\tau}), \tag{26}$$

where $\Delta\tau = \frac{\Delta T}{P}$ and $\langle |\Delta\vec{\rho}|^2 \rangle = \frac{MSD}{\left(\frac{2DP}{1-S}\right)}$ are non-dimensional quantities. Eqs. (17) and (23) link these scales to the original model parameters. Eq. (26) validates the choices we made for the numerical solution, discussed in Section 3. Figure 4 plots $\langle |\Delta\vec{\rho}|^2 \rangle$ versus $\Delta\tau$ for different values of $S$: the larger $S$, the larger the value of $\Delta\tau$ for which the short-time behavior is diffusive.

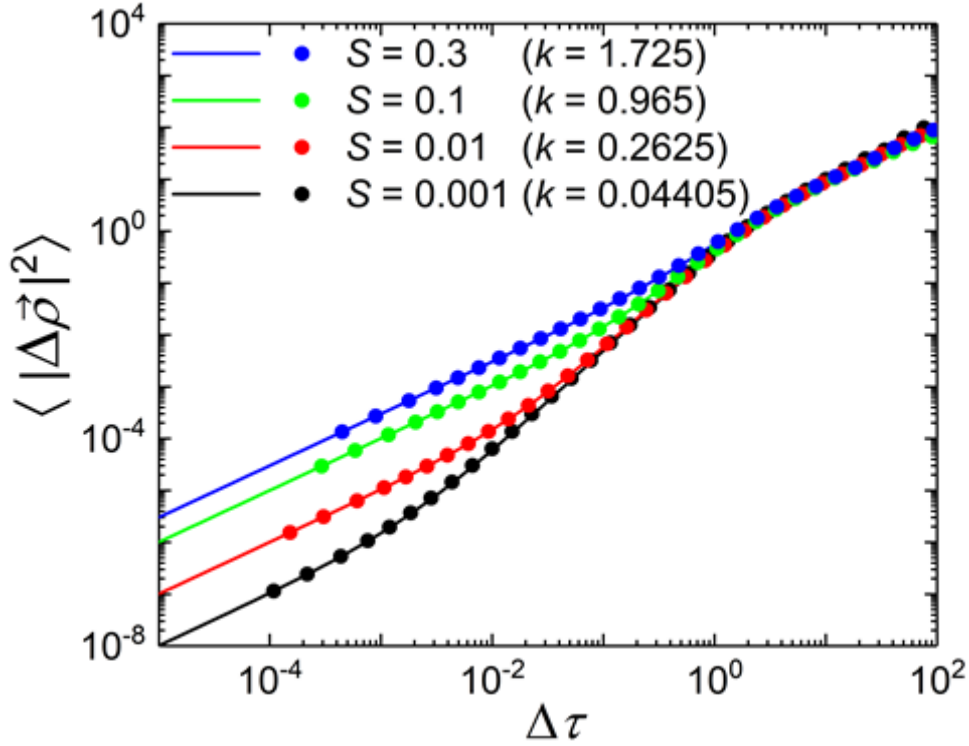

Figure 5. Log-log plot of $\langle |\Delta\vec{\rho}|^2 \rangle$ versus $\Delta\tau$ for q = 0.1, g = 10, γ = 1, and four values of S (S ∈ {0.001, 0.01, 0.1, 0.3}) corresponding to four values of k (k ∈ {0.04405, 0.2625, 0.965, 1.7425}). Solid lines correspond to Eq. (22), while the dots are averages over 100 independent numerical trajectories. Error bars for the simulations are smaller than the dot size.

### 3.3 Analytical forms for the Velocity Auto-Correlation Functions

The diffusive behavior of the position at short-time intervals for $S > 0$ implies that the instantaneous velocity diverges. The instantaneous velocity in natural units, $\vec{u}(\tau)$, is:

$$\vec{u}(\tau) = \lim_{\delta \to 0} \frac{\vec{\rho}(\tau+\delta) - \vec{\rho}(\tau)}{\delta} = \lim_{\delta \to 0} \frac{\Delta\vec{\rho}_\parallel + \Delta\vec{\rho}_\perp}{\delta} = u_\parallel(t)\,\vec{p}(t) + \lim_{\delta \to 0} \frac{\Delta\rho_\perp}{\delta}\,\vec{n}(t), \tag{27}$$

where $\Delta\vec{\rho}_\parallel$ and $\Delta\vec{\rho}_\perp$ are non-dimensional displacements respectively parallel and orthogonal to the polarization. When $k > 0$ and $q > 0$, displacement in the orthogonal direction, $\lim_{\delta \to 0} \frac{\Delta\rho_\perp}{\delta}$ goes to infinity, since $\Delta\rho_\perp$ follows a Wiener process, while $u_\parallel(\tau)$ is well-defined. An experiment cannot always measure $\Delta\vec{\rho}_\parallel$ and $\Delta\vec{\rho}_\perp$ separately. Below, we define two different correlation functions, which account for finite time precision explicitly.

To analyze the divergence of the instantaneous speed $|\vec{u}(\tau)|$, we define the mean velocity over a finite time interval $\delta$:

$$\overline{\vec{u}(\tau,\delta)} \equiv \frac{\vec{p}(\tau+\delta) - \vec{p}(\tau)}{\delta}. \quad (28)$$

Figure 5 shows the *mean speed* $\langle|\overline{\vec{u}(\tau,\delta)}|\rangle$ vs $\delta$ for numerical calculations: the mean speed $\langle|\overline{\vec{u}(\tau,\delta)}|\rangle$ diverges as $\delta \to 0$.

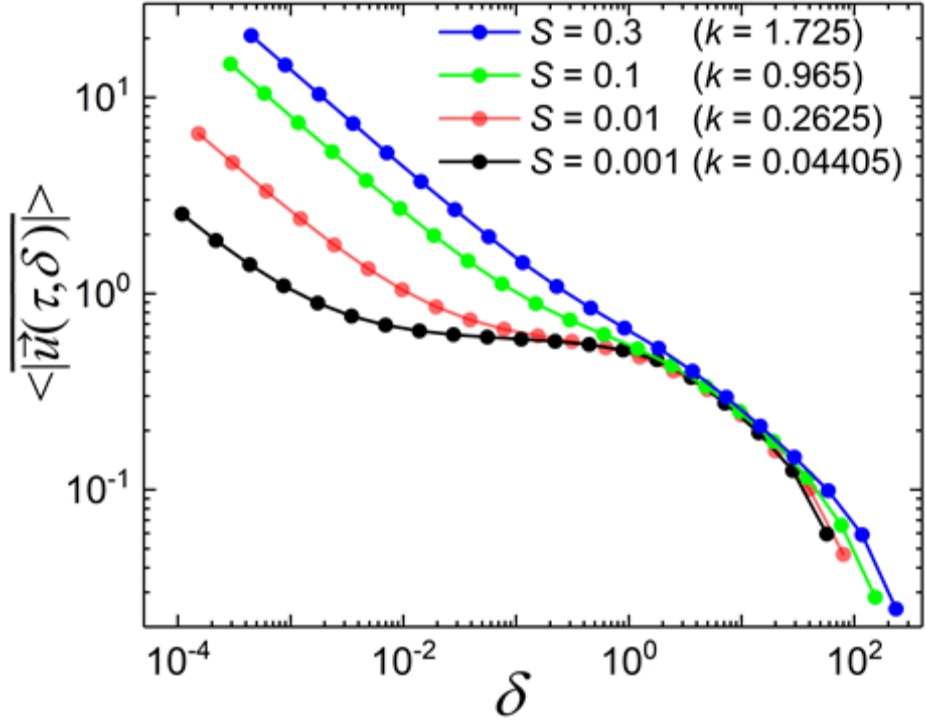

Figure 6. Log-log plot, with time and length rescaled by $P$ and $\sqrt{\frac{2DP}{1-S}}$, of the average mean speed $\langle|\overline{\vec{u}(\tau,\delta)}|\rangle$, obtained by averaging 100 replicas of numerical trajectories, as a function of the time interval $\delta$, for $q = 0.1$, $g = 10$, $\gamma = 1$, and four values of $S$ ($S \in \{0.001, 0.01, 0.1, 0.3\}$) corresponding to four values of $k$ ($k \in \{0.04405, 0.2625, 0.965, 1.7425\}$). Note that $\langle|\overline{\vec{u}(\tau,\delta)}|\rangle$ diverges as $\delta \to 0$.

### 4.3.1 Analytical forms for the Langevin Velocity Auto-Correlation Function: *VACF*

We first observe that

$$\left\langle \left[ v_\parallel(T)\vec{p}(T) + \lim_{\delta\to 0}\frac{\Delta r_\perp(T)}{\delta}\vec{n}(T) \right] \cdot \left[ v_\parallel(T+\Delta T)\vec{p}(T+\Delta T) + \lim_{\delta\to 0}\frac{\Delta r_\perp(T+\Delta T)}{\delta}\vec{n}(T+\Delta T) \right] \right\rangle$$
$$= \langle v_\parallel(T)\vec{p}(T) \cdot v_\parallel(T+\Delta T)\vec{p}(T+\Delta T) \rangle, \quad (29)$$

because $\Delta r_\perp(T)$ obeys a Wiener process with zero average.

We define $VACF$ to be:

$$VACF(\Delta T) = \langle v_\parallel(T)\vec{p}(T) \cdot v_\parallel(T+\Delta T)\vec{p}(T+\Delta T) \rangle. \quad (30)$$

We partition the finite time interval $\Delta T = n\,\Delta t$ into an infinite number $n$ of infinitesimal time intervals $\Delta t$ (such that $\Delta T$ remains finite), sum over it and find (see the Supplementary Materials Online):

$$VACF(\Delta T) = \langle v_\parallel{}^2(T)\rangle e^{-(\gamma+2k)\Delta T} = \langle v_{\parallel sta}{}^2\rangle e^{-\Delta T/P}, \tag{31}$$

as expected. As the asymptotic solution is stationary, $VACF(\Delta T)$ is equal to half the second derivative of the *MSD* curve. Since this second derivative is the same for both Eqs. (21) (Fürth *MSD*) and (22) (modified Fürth *MSD*), the *VACF* has the same form for both models. The result is an exponential decay with a decay constant given by $P$ (and independent of $R$).

### 4.3.2 Mean Velocity Auto-Correlation Function $\psi(\delta,\Delta T)$: Effect of finite-precision measurements

Eq. (31) implies that in the stationary state, $\lim_{\Delta T \to 0} VACF(\Delta T) = \langle v_{\parallel sta}{}^2\rangle$. Experiments and simulations often deviate from Eq. (31), due to two different effects, which we discuss below.

#### 4.3.2.1 Instantaneous velocity is ill-defined for Wiener displacements of position

The definition of instantaneous velocity (Eq. (27)) agrees with the experimental and computational procedure for estimating $\vec{u}$. We measure displacements over time intervals $\delta$ and take the limit of the ratio as $\delta \to 0$: $\vec{u} = \lim_{\delta \to 0} \frac{\Delta \vec{\rho}(\delta)}{\delta} = \lim_{\delta \to 0}\left[\frac{\Delta\rho_\parallel(\delta)}{\delta}\vec{p} + \frac{\Delta\rho_\perp(\delta)}{\delta}\vec{n}\right]$. For a Wiener process for the position, the ratio $\frac{\Delta\rho_\perp(\delta)}{\delta}$ diverges as $\delta \to 0$, so the velocity diverges.

However, in experiments and simulations the limit $\delta \to 0$ is not taken and velocity is estimated using a finite value for $\delta$. When $\delta > S$, the measured particle displacement is in the intermediate-time-interval regime, meaning that the particle movement is ballistic and $u_\parallel \delta \gg \Delta\rho_\perp$. In this case, $\vec{u}(\tau) \approx u_\parallel(\tau)\vec{p}(\tau)$ and estimating the *VACF* using $u_\parallel(\tau)\vec{p}(\tau)$ instead of $\vec{u}(\tau)$ will agree with the prediction of Eq. (31), so the instantaneous velocity is effectively well-defined.

On the other hand, when $\delta$ is finite but $\delta < S$, the second term on the right-hand side of Eq. (27) dominates and the estimated value for the velocity is $\vec{u}(\tau) \approx \frac{\Delta\rho_\perp(\tau)}{\delta}\vec{n}(\tau)$, yielding an estimate of the *VACF* that goes to zero for decreasing $\Delta\tau$, since $\Delta\rho_\perp$ follows a Wiener process.

Here, we use the mean velocity calculated at a finite interval $\delta$ to define the *dimensionless mean velocity autocorrelation function $\psi(\delta,\Delta\tau)$* (with time and length rescaled to be non-dimensional using their natural scales):

$$\psi(\delta,\Delta\tau) \equiv \langle \overline{\vec{u}(\tau,\delta)} \cdot \overline{\vec{u}(\tau+\Delta\tau,\delta)} \rangle, \tag{32}$$

where $\overline{\vec{u}(\tau,\delta)} = \frac{\Delta\rho_\parallel(\delta)}{\delta}\vec{p} + \frac{\Delta\rho_\perp(\delta)}{\delta}\vec{n}$. For infinite-precision measurements we trivially find:

$$\psi(\delta,\Delta\tau) = \frac{(\gamma+2k)}{\gamma}\frac{\left(1-e^{-\gamma\delta/(\gamma+2k)}\right)\left(1-e^{-\delta}\right)}{\delta^2}\langle u_{\parallel sta}{}^2\rangle e^{-\Delta\tau}. \tag{33}$$

For high-precision measurements, small values of $\delta$ imply $\psi(\delta,\Delta\tau) \sim \langle u_{\parallel sta}{}^2\rangle e^{-\Delta\tau}$; that is, $\psi(\delta,\Delta\tau)$ tends to $VACF(\Delta\tau)$. For finite-precision measurements, however, $\psi(\delta,\Delta\tau)$ decreases with decreasing $\Delta\tau$, when $\Delta\tau < S$, due to the poor estimate of $\langle u_{\parallel sta}{}^2\rangle$. If we degrade the precision of our estimate of the mean velocity by truncating the estimate to a fixed number of decimal digits, we see that $\psi(\delta,\Delta\tau)$ decreases as $\Delta\tau$ decreases (Figure 6).

We observe that anisotropy is a necessary condition for predicting that when $\delta \to 0$, $\overline{\vec{u}(\tau,\delta)} \cdot \vec{p}$ converges, while the component orthogonal to the polarization diverges.

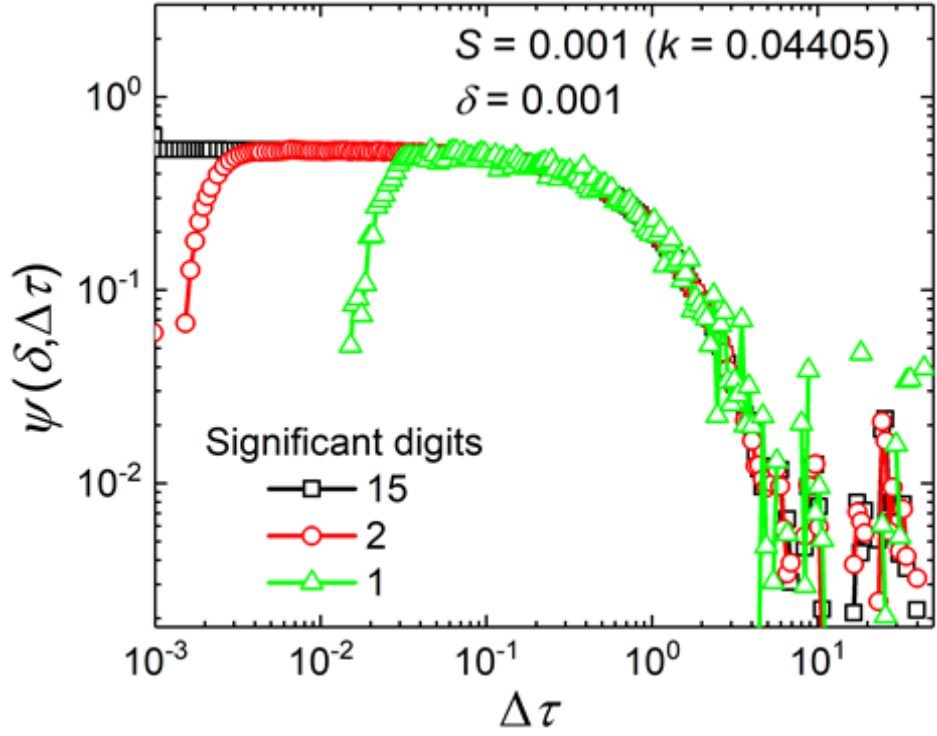

*Figure 7. Log-log plot of $\psi(\delta, \Delta\tau)$ versus $\Delta\tau$, with time and distance rescaled in natural units for q = 0.1, g = 10, γ = 1, k = 0.04405 (S = 0.001), for δ = 0.001 and different precision for the calculation of the mean displacement. For lower precision, estimates of position or velocity $\psi(\delta, \Delta\tau)$ decrease as $\Delta\tau$ decreases.*

### 3.3.2.2 Excessively short time intervals $\Delta\tau$.

Since $\delta$ is not infinitesimal, we must guarantee that $\Delta\tau > \delta$ to prevent the time intervals $[\tau, \tau + \delta]$ and $[\tau + \Delta\tau, \tau + \Delta\tau + \delta]$ from overlapping. Since we use these intervals to estimate, respectively, $\overline{\vec{u}(\tau, \delta)}$ and $\overline{\vec{u}(\tau + \Delta\tau, \delta)}$, when $\Delta\tau < \delta$, the overlap of time intervals introduces a correlation between the displacements used to calculate these quantities. This spurious correlation happens even when the accuracy of measurement is high (Figure 8). For low-precision measurements of displacement and $\Delta\tau < \delta$ (not shown), as $\Delta\tau$ decreases $\psi(\delta, \Delta\tau)$ may first decrease, then increase back to $\psi(\delta, \Delta\tau = 0) = \langle \overline{\vec{u}(\tau, \delta)}^2 \rangle$, which is its maximum value.

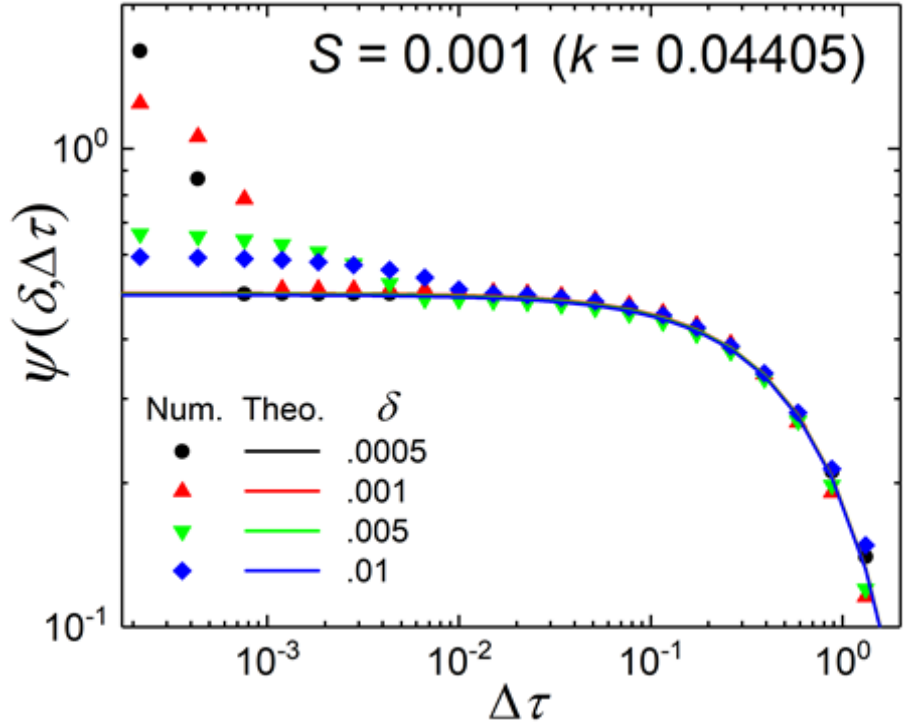

*Figure 8. Log-log plot of $\psi(\delta,\Delta\tau)$ versus $\Delta\tau$, with time and distance measured in natural units for q = 0.1, g = 10, $\gamma$ = 1, k = 0.04405 (S = 0.001), for different values of $\delta > \Delta\tau$. Solid lines correspond to analytical calculations using Eq. (33), and dots correspond to numerical solutions averaged over 10 trajectories. $\delta$ values indicated in the figure.*

**5 Discussion and conclusions**

Migrating cells are anisotropic and their speed is persistent. In their original form, Ornstein-Uhlenbeck processes stem from *isotropic* Langevin models with well-defined instantaneous velocities. Active-Matter models are anisotropic. In models where the cell speed follows some dynamics in one direction and cell displacement in the orthogonal directions obeys a Wiener process, instantaneous velocity is ill-defined. Not surprisingly, Active-Matter models generally avoid dynamical equations for velocity, assuming overdamped particles. However, both Ornstein-Uhlenbeck and Active-Matter models can successfully describe the movement of migrating cells on flat surfaces, so in appropriate limits, they must yield the same observable results.

We proposed and solved an Anisotropic Ornstein-Uhlenbeck process for a particle with a time-varying internal polarization, with a well-defined instantaneous velocity in the direction of polarization, and with a pure Brownian motion in the direction orthogonal to the polarization. This model couples a Langevin equation for velocity in the polarization direction to a Wiener process for displacements in the direction perpendicular to polarization and an Ornstein-Uhlenbeck process for the evolution of the direction of polarization. The main results are: i) analytically-derived expressions agree with empirical *MSD* and *VACF* curves obtained for experiments and CompuCell3D simulations, ii) *MSD* curves show a diffusive regime for short-time intervals as in experiments and simulations, iii) procedures that give meaningful estimates for the *MSD* and *VACF* curves despite finite-precision measurements of speed and velocity, and iv) the definition of time and length scales (as in Ref. [9]), that enable comparison of movement statistics between experiments and between experiments and simulations.

We previously used Eq. (22) to fit 12 different sets of migrating cell experiments, from 5 different laboratories [9], as well as CompuCell3D simulations of migrating cells [10]. The observed behaviors of the *MSD*, speed, and velocity autocorrelation functions in these experiments and simulations agree with our analytical calculations. Recent CompuCell3D simulations, proposing a measure for cell polarization, showed that velocities parallel and orthogonal to polarization behave differently and indicate that anisotropy should be considered in any analysis of cell-migration statistics [23]. The specific functional forms of our prediction remain to be verified in cell-tracking experiments.

This statistical analysis allows quantification of particle trajectories in Active-Matter and biological models or obtained from biological experiments, as long as their movement obeys the Anisotropic Ornstein-Uhlenbeck process defined in Equations 4-10.

**Acknowledgements**

This work has been partially financed by Brazilian Agencies CNPq, CAPES (grant # 0001), and FAPERGS. J.A.G. acknowledges support from US National Institutes of Health grants U24 EB028887 and R01 GM122424 and National Science Foundation grant NSF 1720625.

# Supplementary Materials for
# Exact solution for the anisotropic Ornstein-Uhlenbeck Process

Rita M.C. de Almeida[1,2,3], Guilherme S. Y. Giardini[1], Mendeli Vainstein[1],
James A. Glazier[4], and Gilberto L. Thomas[1]

[1] Instituto de Física, Universidade Federal do Rio Grande do Sul, Porto Alegre, RS, Brazil
[2] Instituto Nacional de Ciência e Tecnologia: Sistemas Complexos, Universidade Federal do Rio Grande do Sul, Porto Alegre, RS, Brazil
[3] Programa de Pós Graduação em Bioinformática, Universidade Federal do Rio Grande do Norte, Natal, RN, Brazil
[4] Biocomplexity Institute and Department of Intelligent Systems Engineering, Indiana University, Bloomington, Indiana, United States of America

April 21, 2021

## Analytic solutions

### Velocity in the instantaneous polarization direction

We start at $t = 0$ with the speed $v_{\|0}$ (which can be either positive or negative) in the direction of the cell polarization $\vec{p}_0$. Using Eq. (5) in the main text, we can write the velocity in the direction of cell polarization at the subsequent small time interval $\Delta t$ as:

$$v_{\|}(\Delta t)\vec{p}(\Delta t) = \left[ (1 - \gamma \Delta t)\, v_{\|0} + \int_0^{\Delta t} \xi_{\|}(t)\, \mathrm{d}t \right] (\vec{p}(0) \cdot \vec{p}(\Delta t))\, \vec{p}(\Delta t). \qquad (1)$$

After $n$ iterations, the velocity at time $T = n\,\Delta t$ is:

$$\begin{aligned}
v_{\|}(n\,\Delta t)\vec{p}(n\,\Delta t) &= v_{\|0}\,(1 - \gamma\Delta t)^n\,[\vec{p}_0 \cdot \vec{p}(\Delta t)]\,[\vec{p}(\Delta t) \cdot \vec{p}(2\,\Delta t)]\ldots[\vec{p}((n-1)\Delta t) \cdot \vec{p}(n\Delta t)]\,\vec{p}(n\Delta t) \\
&\quad + \int_0^{\Delta t} \mathrm{d}s\, \xi_{\|}(s)\,(1 - \gamma\Delta t)^{n-1}\,[\vec{p}_0 \cdot \vec{p}(\Delta t)]\ldots[\vec{p}((n-1)\Delta t) \cdot \vec{p}(n\Delta t)]\,\vec{p}(n\Delta t) \\
&\quad + \int_{\Delta t}^{2\Delta t} \mathrm{d}s\, \xi_{\|}(s)\,(1 - \gamma\Delta t)^{n-2}\,[\vec{p}(\Delta t) \cdot \vec{p}(2\Delta t)]\ldots[\vec{p}((n-1)\Delta t) \cdot \vec{p}(n\Delta t)]\,\vec{p}(n\Delta t) \\
&\quad + \ldots \\
&\quad + \int_{(n-1)\Delta t}^{n\Delta t} \mathrm{d}s\, \xi_{\|}(s)\,[\vec{p}((n-1)\Delta t) \cdot \vec{p}(n\Delta t)]\,\vec{p}(n\Delta t). \qquad (2)
\end{aligned}$$

Observe that at each time step the realignment of the polarization axis implies that only the component of the velocity in the direction of the new polarization is partially conserved. We will calculate each of the terms in the above equations separately. First we need to deal with the scalar products between cell polarizations at different time instants.

Cell polarization vectors are unitary. Hence $\vec{p}_i \cdot \vec{p}_j = \cos(\theta_i - \theta_j)$, where $\theta_j$ is the angle between the direction of $\vec{p}_j$ and the abscissas axis. We define $\Delta\theta_{i,j} = \theta_i - \theta_j$.

We assume that the change in polarization direction is due to the noise perpendicular to the instantaneous polarization direction. This noise is also responsible for displacements in the perpendicular direction. Hence, for a small time interval $\Delta t$, we assume that the change in direction is a Wiener process, such that $\langle (\Delta\theta_{i,i+1})^2 \rangle_{\beta_\perp} = 2k\Delta t$. Observe also that $\xi_{\|}$ and $\beta_\perp$ are not correlated, hence the average over the two noises decouples.

We are interested in the $\Delta t \to 0$ limit, which means that the scalar product between consecutive polarization vectors may be taken up to first order in $\Delta t$. In this case, the following approximations apply:

$$\cos \Delta\theta_{i,i+1} \approx 1 - \frac{(\Delta\theta_{i,i+1})^2}{2}, \qquad (3)$$

$$\cos^2 \Delta\theta_{i,i+1} \approx 1 - (\Delta\theta_{i,i+1})^2. \qquad (4)$$

Also, for any pair $(i, j)$, with $i < j$ we have:

$$\Delta \theta_{i,j} = \Delta \theta_{i,i+1} + \Delta \theta_{i+1,i+2} + \ldots + \Delta \theta_{j-1,j}, \tag{5}$$

and hence:

$$\begin{aligned}
\langle \cos \Delta \theta_{i,j} \rangle_{\beta_\perp} &= \langle \cos \Delta \theta_{i,i+1} \rangle_{\beta_\perp} \langle \cos(\Delta \theta_{i+1,i+2} + \Delta \theta_{i+2,i+3} + \ldots + \Delta \theta_{j-1,j}) \rangle_{\beta_\perp} \\
&\quad - \langle \sin \Delta \theta_{i,i+1} \rangle_{\beta_\perp} \langle \sin(\Delta \theta_{i+1,i+2} + \Delta \theta_{i+2,i+3} + \ldots + \Delta \theta_{j-1,j}) \rangle_{\beta_\perp} \\
&\approx \langle \cos \Delta \theta_{i,i+1} \rangle_{\beta_\perp} \langle \cos(\Delta \theta_{i+1,i+2} + \Delta \theta_{i+2,i+3} + \ldots + \Delta \theta_{j-1,j}) \rangle_{\beta_\perp} \\
&\approx \langle \cos \Delta \theta_{i,i+1} \rangle_{\beta_\perp} \langle \cos \Delta \theta_{i+1,i+2} \rangle_{\beta_\perp} \ldots \langle \cos \Delta \theta_{j-1,j} \rangle_{\beta_\perp} \\
&\approx (1 - k \Delta t)^{|j-i|}.
\end{aligned} \tag{6}$$

Also, in the limit $\Delta t \to \infty$,

$$\langle \cos^2 \Delta \theta_{i,j} \rangle_{\beta_\perp} \approx (1 - k \Delta t)^{2|j-i|}. \tag{7}$$

With this approximation, we can calculate $<v_\parallel>$ from Eq. (2) above. Since the terms containing noise averages vanish, we have:

$$<v_\parallel> = v_{\parallel 0} (1 - \gamma \Delta t)^n (1 - k \Delta t)^n, \tag{8}$$

which, using $T = n \Delta t$, and taking the limit $n \to \infty$, can be written as:

$$<v_\parallel> = v_{\parallel 0} \exp\left[-(\gamma + k)T\right]. \tag{9}$$

So, for $T \to \infty$,

$$<v_\parallel> = 0. \tag{10}$$

Now, we square the velocity, Eq. (2), and take the average over the noise. We get:

$$\begin{aligned}
\langle v_\parallel^2(n\Delta t) \rangle &= v_{\parallel 0}^2 (1 - \gamma \Delta t)^{2n} \langle [\vec{p}_0 \cdot \vec{p}(\Delta t)]^2 [\vec{p}(\Delta t) \cdot \vec{p}(2\Delta t)]^2 \ldots [\vec{p}((n-1)\Delta t) \cdot \vec{p}(n\Delta t)]^2 \rangle \\
&\quad + g \int_0^{\Delta t} \mathrm{d}s \, (1 - \gamma \Delta t)^{2(n-1)} \langle [\vec{p}_0 \cdot \vec{p}(\Delta t)]^2 \ldots [\vec{p}((n-1)\Delta t) \cdot \vec{p}(n\Delta t)]^2 \rangle \\
&\quad + g \int_{\Delta t}^{2\Delta t} \mathrm{d}s \, (1 - \gamma \Delta t)^{2(n-2)} \langle [\vec{p}(\Delta t) \cdot \vec{p}(2\Delta t)]^2 \ldots [\vec{p}((n-1)\Delta t) \cdot \vec{p}(n\Delta t)]^2 \rangle \\
&\quad + \ldots \\
&\quad + g \int_{(n-1)\Delta t}^{n\Delta t} \mathrm{d}s \, \langle [\vec{p}((n-1)\Delta t) \cdot \vec{p}(n\Delta t)]^2 \rangle,
\end{aligned} \tag{11}$$

where we have used:

$$\langle \xi_\parallel(t) \xi_\parallel(t') \rangle = g \delta(t - t'), \tag{12}$$

and noted that cross terms with different integral limits vanish.

As polarization vectors are unitary, we also use $[\vec{p}((j-1)\Delta t) \cdot \vec{p}(j\Delta t)] = \cos(\theta((j-1)\Delta t) - \theta(j\Delta t)) = \cos(\Delta \theta)$, which, in turn, for small $\Delta t$, approaches $\cos(\Delta \theta) \sim 1 - \frac{1}{2}(\Delta \theta)^2$ and $\cos^2(\Delta \theta) \sim 1 - (\Delta \theta)^2$. Now, assuming that the change in polarization direction is due to the Wiener process perpendicular to the polarization axis, averaging over the perpendicular noise in each time interval of duration $\Delta t$, leads to $\langle (\Delta \theta)^2 \rangle = 2k\Delta t$. Using this result in Eq. (11), and solving the integrals, we find:

$$\begin{aligned}
\langle v_\parallel^2(n\Delta t) \rangle &= v_{\parallel 0}^2 (1 - \gamma \Delta t)^{2n} (1 - k \Delta t)^{2n} \\
&\quad + g \left[(1 - \gamma \Delta t)^{2(n-1)} (1 - k \Delta t)^{2(n-1)} + \ldots + 1 \right].
\end{aligned} \tag{13}$$

Since $T = n\Delta t$ and taking the limits $\Delta t \to 0$ and $n \to \infty$, such that $T$ is finite, we obtain:

$$\langle v_\parallel^2(T) \rangle = \frac{g}{2(\gamma + k)} + \left(v_{\parallel 0}^2 - \frac{g}{2(\gamma + k)}\right) \exp\left[-2(\gamma + k)T\right]. \tag{14}$$

Making $v_{\parallel 0}^2 = \frac{g}{2(\gamma + k)}$ from the start eliminates any transients.

## Mean-squared displacement

We obtain the mean-squared displacement ($MSD$) by first calculating the displacement in each time interval $\Delta t$, from $t = 0$ to $t = n\Delta t = \Delta T$, then summing over time, taking the square of this expression and finally averaging over noise. In this section we will simplify our notation by writing $\vec{p}(n\Delta t) = \vec{p}_n$.

**Displacement in each time interval**

At $t = 0$, we assume the cell has a polarization $\vec{p}_0$ that remains constant during the next small time interval $\Delta t$. The cell displacement at the end of this interval $\Delta t$ is, then,

$$\vec{r}(\Delta t) - \vec{r}(0) = v_{\|0} \Delta t\, \vec{p}_0 + \int_0^{\Delta t} \mathsf{d}s\, (\Delta t - s)\, \xi_\|(s)\, \vec{p}_0 + \int_0^{\Delta t} \mathsf{d}s\, \xi_\perp(s)\, \vec{n}_0\,. \qquad (15)$$

At $t = \Delta t$, the cell polarization changes, going from $\vec{p}_0$ to $\vec{p}_1 = \vec{p}(\Delta t)$. So, for the following time interval, the initial value of the speed is $v_\|(\Delta t)\vec{p}_0 \cdot \vec{p}_1$. Therefore, the next displacement is

$$\vec{r}(2\Delta t) - \vec{r}(\Delta t) = \Delta t\, (\vec{v_\|}(\Delta t) \cdot \vec{p}_1)\vec{p}_1 + \int_{\Delta t}^{2\Delta t} \mathsf{d}s\, (\Delta t - s)\, \xi_\|(s)\, \vec{p}_1 + \int_{\Delta t}^{2\Delta t} \mathsf{d}s\, \xi_\perp(s)\, \vec{n}_1\,. \qquad (16)$$

Since,

$$\vec{v_\|}(\Delta t) = (1 - \gamma \Delta t)\, v_{\|0}\, \vec{p}_0 + \int_0^{\Delta t} \mathsf{d}s\, \xi_\|(s)\, \vec{p}_0\,, \qquad (17)$$

we obtain:

$$\begin{aligned}\vec{r}(2\Delta t) - \vec{r}(\Delta t) &= (1 - \gamma \Delta t)\, \Delta t\, v_{\|0}\, (\vec{p}_0 \cdot \vec{p}_1)\, \vec{p}_1 + \Delta t \int_0^{\Delta t} \mathsf{d}s\, \xi_\|(s)\, (\vec{p}_0 \cdot \vec{p}_1)\, \vec{p}_1 \\ &\quad + \int_{\Delta t}^{2\Delta t} \mathsf{d}s\, (\Delta t - s)\, \xi_\|(s)\, \vec{p}_1 + \int_{\Delta t}^{2\Delta t} \mathsf{d}s\, \xi_\perp(s)\, \vec{n}_1\,.\end{aligned} \qquad (18)$$

Generalizing for the $n^{th}$ time interval,

$$\begin{aligned}\vec{r}(n\Delta t) - \vec{r}((n-1)\Delta t) &= (1 - \gamma\Delta t)^{n-1}\, \Delta t\, v_{\|0}\, (\vec{p}_0 \cdot \vec{p}_1)\, (\vec{p}_1 \cdot \vec{p}_2) \ldots (\vec{p}_{n-2} \cdot \vec{p}_{n-1})\, \vec{p}_{n-1} \\ &\quad + \Delta t \int_0^{\Delta t} \mathsf{d}s\, \xi_\|(s)\, (1-\gamma\Delta t)^{n-2}\, (\vec{p}_0 \cdot \vec{p}_1)(\vec{p}_1 \cdot \vec{p}_2)\ldots(\vec{p}_{n-2} \cdot \vec{p}_{n-1})\, \vec{p}_{n-1} \\ &\quad + \Delta t \int_{\Delta t}^{2\Delta t} \mathsf{d}s\, \xi_\|(s)\, (1-\gamma\Delta t)^{n-3}\, (\vec{p}_1 \cdot \vec{p}_2)(\vec{p}_2 \cdot \vec{p}_3)\ldots(\vec{p}_{n-2} \cdot \vec{p}_{n-1})\, \vec{p}_{n-1} \\ &\quad \ldots \\ &\quad + \Delta t \int_{(n-2)\Delta t}^{(n-1)\Delta t} \mathsf{d}s\, \xi_\|(s)\, (\vec{p}_{n-2} \cdot \vec{p}_{n-1})\, \vec{p}_{n-1} \\ &\quad + \int_{(n-1)\Delta t}^{n\Delta t} \mathsf{d}s\, (\Delta t - s)\, \xi_\|(s)\, \vec{p}_{n-1} \\ &\quad + \int_{(n-1)\Delta t}^{n\Delta t} \mathsf{d}s\, \xi_\perp(s)\, \vec{n}_{n-1}\,.\end{aligned} \qquad (19)$$

We obtain the total displacement, from $t = 0$ to $t = n\Delta t$, by summing over the individual time steps, *i.e.*,

$$\begin{aligned}
\vec{r}(n\Delta t) - \vec{r}(0) = v_{\|0}\Delta t &\left[(1-\gamma\Delta t)^{n-1}(\vec{p}_0 \cdot \vec{p}_1)(\vec{p}_1 \cdot \vec{p}_2)\ldots(\vec{p}_{n-2} \cdot \vec{p}_{n-1})\vec{p}_{n-1}\right), \cdot \\
&+ (1-\gamma\Delta t)^{n-2}(\vec{p}_0 \cdot \vec{p}_1)(\vec{p}_1 \cdot \vec{p}_2)\ldots(\vec{p}_{n-3} \cdot \vec{p}_{n-2})\vec{p}_{n-2} \\
&+ \ldots \\
&+ \vec{p}_0 \Big] \\
+ \Delta t \int_0^{\Delta t} &\mathsf{d}s\, \xi_\|(s) \Big[(1-\gamma\Delta t)^{n-2}(\vec{p}_0 \cdot \vec{p}_1)(\vec{p}_1 \cdot \vec{p}_2)\ldots(\vec{p}_{n-2} \cdot \vec{p}_{n-1})\vec{p}_{n-1} \\
&+ (1-\gamma\Delta t)^{n-3}(\vec{p}_0 \cdot \vec{p}_1)(\vec{p}_1 \cdot \vec{p}_2)\ldots(\vec{p}_{n-3} \cdot \vec{p}_{n-2})\vec{p}_{n-2} \\
&+ \ldots \\
&+ (\vec{p}_0 \cdot \vec{p}_1)\vec{p}_1 \Big] \\
+ \int_0^{\Delta t} &\mathsf{d}s\, \xi_\|(s)(\Delta t - s)\vec{p}_0 \\
+ \Delta t \int_{\Delta t}^{2\Delta t} &\mathsf{d}s\, \xi_\|(s) \Big[(1-\gamma\Delta t)^{n-3}(\vec{p}_1 \cdot \vec{p}_2)(\vec{p}_2 \cdot \vec{p}_3)\ldots(\vec{p}_{n-2} \cdot \vec{p}_{n-1})\vec{p}_{n-1} \\
&+ (1-\gamma\Delta t)^{n-4}(\vec{p}_1 \cdot \vec{p}_2)(\vec{p}_2 \cdot \vec{p}_3)\ldots(\vec{p}_{n-3} \cdot \vec{p}_{n-2})\vec{p}_{n-2} \\
&+ \ldots \\
&+ (\vec{p}_1 \cdot \vec{p}_2)\vec{p}_2 \Big] \\
+ \int_{\Delta t}^{2\Delta t} &\mathsf{d}s\, \xi_\|(s)(\Delta t_2 - s)\vec{p}_1 \\
+ \ldots & \\
+ \Delta t \int_{(n-2)\Delta t}^{(n-1)\Delta t} &\mathsf{d}s\, \xi_\|(s)(\vec{p}_{n-2} \cdot \vec{p}_{n-1})\vec{p}_{n-1} \\
+ \int_{(n-2)\Delta t}^{(n-1)\Delta t} &\mathsf{d}s\, \xi_\|(s)[\Delta t_{n-1} - s]\vec{p}_{n-2} \\
+ \int_{(n-1)\Delta t}^{(n)\Delta t} &\mathsf{d}s\, \xi_\|(s)(\Delta t_n - s)\vec{p}_{n-1} \\
+ \int_0^{\Delta t} &\mathsf{d}s\, \xi_\perp(s)\vec{n}_0 + \ldots + \int_{(n-1)\Delta t}^{n\Delta t} \mathsf{d}s\, \xi_\perp(s)\vec{n}_{n-1}\,.
\end{aligned} \tag{20}$$

**Squared displacement and average over noises**

To obtain the $MSD$ we must square Eq. (20) and then average over the noise. In this process, the square of each term is added to the cross terms. However, some cross terms vanish due to the noise average. The terms which vanish are: (i) cross terms containing a term that depends on $v_{\|0}$; (ii) cross terms with products of terms integrated over different time intervals; and (iii) terms crossing $\xi_\|(s)$ with $\beta_\perp(s)$. After eliminating these terms, we define the following quantities:

$$\begin{aligned}
I^2 = v_{\|0}^2 (\Delta t)^2 \Big\langle &\Big[(1-\gamma\Delta t)^{n-1}(\vec{p}_0 \cdot \vec{p}_1)(\vec{p}_1 \cdot \vec{p}_2)\ldots(\vec{p}_{n-2} \cdot \vec{p}_{n-1})\vec{p}_{n-1} \\
&+ (1-\gamma\Delta t)^{n-2}(\vec{p}_0 \cdot \vec{p}_1)(\vec{p}_1 \cdot \vec{p}_2)\ldots(\vec{p}_{n-3} \cdot \vec{p}_{n-2})\vec{p}_{n-2} + \ldots \\
&+ (1-\gamma\Delta t)(\vec{p}_0 \cdot \vec{p}_1)\vec{p}_1 + \vec{p}_0 \Big]^2 \Big\rangle_{\xi_\|,\beta_\perp},
\end{aligned} \tag{21}$$

$$\begin{aligned}
J^2(k\Delta t) = \Big\langle \Big\{\Delta t \int_{(k-1)\Delta t}^{k\Delta t} &\mathsf{d}s\, \xi_\|(s)\Big[(1-\gamma\Delta t)^{n-k-1}(\vec{p}_{k-1} \cdot \vec{p}_k)(\vec{p}_k \cdot \vec{p}_{k+1})\ldots(\vec{p}_{n-2} \cdot \vec{p}_{n-1})\vec{p}_{n-1} \\
&+ (1-\gamma\Delta t)^{n-k-2}(\vec{p}_{k-1} \cdot \vec{p}_k)(\vec{p}_k \cdot \vec{p}_{k+1})\ldots(\vec{p}_{n-3} \cdot \vec{p}_{n-2})\vec{p}_{n-2} \\
&+ \ldots + (\vec{p}_{k-1} \cdot \vec{p}_k)\vec{p}_k\Big] \\
+ \int_{(k-1)\Delta t}^{k\Delta t} &\mathsf{d}s\, \xi_\|(s)(k\Delta t - s)\vec{p}_{k-1}\Big\}^2 \Big\rangle_{\xi_\|,\beta_\perp},
\end{aligned} \tag{22}$$

and,

$$K^2 = \Big\langle \Big[\int_0^{n\Delta t} \mathsf{d}s\, \xi_\perp\Big]^2 \Big\rangle_{\xi_\|,\xi_\perp}, \tag{23}$$

where $\langle \cdot \rangle$ stands for the average over the noise $\xi_\parallel$, in the polarization direction, and $\xi_\perp$, in the direction perpendicular to the polarization direction. With these definitions, we write:

$$\begin{aligned} MSD &= \lim_{\substack{\Delta t \to 0 \\ n \to \infty}} \left\langle |\vec{r}(n\Delta t) - \vec{r}(0)|^2 \right\rangle_{\xi_\parallel, \beta_\perp} \\ &= \lim_{\substack{\Delta t \to 0 \\ n \to \infty}} \left[ I^2 + \sum_{k=1}^{n} J^2(k\Delta t) + K^2 \right]. \end{aligned} \qquad (24)$$

## $I^2$ calculation

In Eq. (21) we write all scalar products $\vec{p}_i \cdot \vec{p}_j$ as cosines:

$$I^2 = v_{\parallel 0}^2 (\Delta t)^2 \left\langle \left[ (1-\gamma\Delta t)^{n-1} \cos\Delta\theta_{0,1} \cos\Delta\theta_{1,2} \ldots \cos\Delta\theta_{n-2,n-1} \vec{p}_{n-1} \right. \right. \\ \left. \left. + (1-\gamma\Delta t)^{n-2} \cos\Delta\theta_{0,1} \cos\Delta\theta_{1,2} \ldots \cos\Delta\theta_{n-3,n-2} \vec{p}_{n-2} + \ldots \right. \right. \\ \left. \left. + (1-\gamma\Delta t) \Delta\theta_{0,1} \vec{p}_1 + \vec{p}_0 \right]^2 \right\rangle_{\xi_\parallel, \beta_\perp}, \qquad (25)$$

which expands into:

$$\begin{aligned} I^2 &= v_{\parallel 0}^2 (\Delta t)^2 \left\langle \left[ \sum_{i=0}^{n-1} (1-\gamma\Delta t)^i \prod_{m=1}^{i} \cos\Delta\theta_{m-1,m} \vec{p}_i \right]^2 \right\rangle_{\xi_\parallel, \beta_\perp} \\ &= v_{\parallel 0}^2 (\Delta t)^2 \left\langle \sum_{i=0}^{n-1} \sum_{j=0}^{n-1} (1-\gamma\Delta t)^{i+j} \prod_{m=1}^{i} \cos\Delta\theta_{m-1,m} \prod_{q=1}^{j} \cos\Delta\theta_{q-1,q} (\vec{p}_i \cdot \vec{p}_j) \right\rangle_{\xi_\parallel, \beta_\perp} \\ &= v_{\parallel 0}^2 (\Delta t)^2 \left\langle \sum_{i=0}^{n-1} (1-\gamma\Delta t)^{2i} \prod_{m=1}^{i} \cos^2\Delta\theta_{m-1,m} \right\rangle_{\xi_\parallel, \beta_\perp} \\ &+ 2 v_{\parallel 0}^2 (\Delta t)^2 \left\langle \sum_{i=0}^{n-2} \sum_{j=i+1}^{n-1} (1-\gamma\Delta t)^{i+j} \prod_{m=1}^{i} \cos\Delta\theta_{m-1,m} \prod_{q=1}^{j} \cos\Delta\theta_{q-1,q} \prod_{p=i}^{j} \cos\Delta\theta_{p-1,p} \right\rangle_{\xi_\parallel, \beta_\perp}. \end{aligned} \qquad (26)$$

The noise at different time intervals is not correlated. Hence, using Eq. (7), we find:

$$\begin{aligned} I^2 &= v_{\parallel 0}^2 (\Delta t)^2 \sum_{i=0}^{n-1} (1-\gamma\Delta t)^{2i} (1-k\Delta t)^{2i} \\ &+ 2 v_{\parallel 0}^2 (\Delta t)^2 \sum_{i=0}^{n-2} \sum_{j=i+1}^{n-1} (1-\gamma\Delta t)^{i+j} (1-k\Delta t)^{2j}. \end{aligned} \qquad (27)$$

These terms are sums of geometric progressions. After some manipulations, and taking the limits $\Delta t \to 0$ and $n \to \infty$, we find:

$$I^2 = \frac{v_{\parallel 0}^2}{(\gamma+2k)(\gamma+k)\gamma} \left[ \gamma + (\gamma+2k) e^{-2(\gamma+k)\Delta T} - 2(\gamma+k) e^{-(\gamma+2k)\Delta T} \right]. \qquad (28)$$

Eq. (14) gives the asymptotic value of the squared velocity $v_{\parallel 0}^2$ which is $\frac{g}{2(\gamma+k)}$. Taking the initial velocity to be this asymptotic value, we can write $I^2$ as:

$$I^2 = \frac{g}{2(\gamma+2k)(\gamma+k)^2 \gamma} \left[ \gamma + (\gamma+2k) e^{-2(\gamma+k)\Delta T} - 2(\gamma+k) e^{-(\gamma+2k)\Delta T} \right]. \qquad (29)$$

Observe that when $k = 0$,

$$I^2(k=0) = \frac{g}{2\gamma^3} \left[ 1 - e^{-(\gamma)\Delta T} \right]^2. \qquad (30)$$

## $J^2$ calculation

Again, in Eq. (22), we write all scalar products $\vec{p}_i \cdot \vec{p}_j$ as cosines:

$$
\begin{aligned}
J^2(k\Delta t) &= \Big\langle \Big\{ \Delta t \int_{(k-1)\Delta t}^{k\Delta t} \mathsf{d}s\, \xi_\|(s) \Big[ (1-\gamma\Delta t)^{n-k-1} \cos\Delta\theta_{k-1,k} \cos\Delta\theta_{k,k+1}\ldots \cos\Delta\theta_{n-2,n-1}\vec{p}_{n-1} \\
&\qquad + (1-\gamma\Delta t)^{n-k-2} \cos\Delta\theta_{k-1,k} \cos\Delta\theta_{k,k+1}\ldots \cos\Delta\theta_{n-3,n-2}\vec{p}_{n-2} \\
&\qquad + \ldots \\
&\qquad + \cos\Delta\theta_{k-1,k}\vec{p}_k \Big] \\
&\quad + \int_{(k-1)\Delta t}^{k\Delta t} \mathsf{d}s\, \xi_\|(s)\, (k\Delta t - s)\,\vec{p}_{k-1} \Big\}^2 \Big\rangle_{\xi_\|,\beta_\perp},
\end{aligned}
\qquad (31)
$$

and expand the square:

$$
\begin{aligned}
J^2(k\Delta t) &= \Big\langle \Big\{ \Delta t \int_{(k-1)\Delta t}^{k\Delta t} \mathsf{d}s\, \xi_\|(s) \Big[ (1-\gamma\Delta t)^{n-k-1} \cos\Delta\theta_{k-1,k} \cos\Delta\theta_{k,k+1}\ldots \cos\Delta\theta_{n-2,n-1}\vec{p}_{n-1} \\
&\qquad + (1-\gamma\Delta t)^{n-k-2} \cos\Delta\theta_{k-1,k} \cos\Delta\theta_{k,k+1}\ldots \cos\Delta\theta_{n-3,n-2}\vec{p}_{n-2} \\
&\qquad + \ldots \\
&\qquad + \cos\Delta\theta_{k-1,k}\vec{p}_k \Big] \Big\}^2 \Big\rangle_{\xi_\|,\beta_\perp} \\
&\quad + \Big\langle \Big\{ \int_{(k-1)\Delta t}^{k\Delta t} \mathsf{d}s\, \xi_\|(s)(k\Delta t - s)\,\vec{p}_{k-1} \Big\}^2 \Big\rangle_{\xi_\|,\beta_\perp} \\
&\quad + 2\Big\langle \Delta t \int_{(k-1)\Delta t}^{k\Delta t} \mathsf{d}s\, \xi_\|(s) \Big[ (1-\gamma\Delta t)^{n-k-1} \cos\Delta\theta_{k-1,k} \cos\Delta\theta_{k,k+1}\ldots \cos\Delta\theta_{n-2,n-1}\vec{p}_{n-1} \\
&\qquad + (1-\gamma\Delta t)^{n-k-2} \cos\Delta\theta_{k-1,k} \cos\Delta\theta_{k,k+1}\ldots \cos\Delta\theta_{n-3,n-2}\vec{p}_{n-2} \\
&\qquad + \ldots \\
&\qquad + \cos\Delta\theta_{k-1,k}\vec{p}_k \Big] \int_{(k-1)\Delta t}^{k\Delta t} \mathsf{d}s\, \xi_\|(s)(k\Delta t - s)\,\vec{p}_{k-1} \Big\rangle_{\xi_\|,\beta_\perp}.
\end{aligned}
\qquad (32)
$$

After long, tedious, but straightforward calculations, the sum over $k$, in the limit of $\Delta t \to 0$, yields:

$$
\begin{aligned}
\sum_{i=0}^{n} J^2(i\Delta t) &= \frac{g}{(\gamma+k)(\gamma+2k)}\Delta T \\
&\quad + \frac{g}{2\gamma(\gamma+k)^2}\left(1 - e^{-2(\gamma+k)\Delta T}\right) \\
&\quad - \frac{2g}{\gamma(\gamma+2k)^2}\left(1 - e^{-(\gamma+2k)\Delta T}\right).
\end{aligned}
\qquad (33)
$$

## $K^2$ calculation

This calculation is rather straightforward:

$$
\langle \xi_\perp(s)\xi_\perp(s')\rangle = q\,\langle \beta_\perp(s)\beta_\perp(s')\rangle = 2\,q\,k\,\delta(s-s'). \qquad (34)
$$

From Eq. (23), the expression for $K^2$ is, then,

$$
K^2 = 2qk\Delta T. \qquad (35)
$$

We present the estimate of $q$ below.

## MSD final expression

Using Eqs. (29), (33), and (35) in Eq. (24), we find:

$$
MSD = \frac{g}{(\gamma+2k)(\gamma+k)}\left[\Delta T - \frac{1}{\gamma+2k}\left(1 - e^{-(\gamma+2k)\Delta T}\right)\right] + 2qk\Delta T, \qquad (36)
$$

which we may rewrite as:

$$
MSD = \frac{g}{(\gamma+2k)^2(\gamma+k)}\left[(1+Q)(\gamma+2k)\Delta T - \left(1 - e^{-(\gamma+2k)\Delta T}\right)\right], \qquad (37)
$$

where,
$$Q = \frac{2qk}{g}(\gamma + 2k)(\gamma + k). \tag{38}$$

Defining the natural time unit $P$ as,
$$P = \frac{1}{(\gamma + 2k)}, \tag{39}$$

we can define the dimensionless quantity $\Delta\tau = \Delta T/P$, and define $S = \frac{Q}{1+Q}$. The $MSD$ becomes:
$$MSD = \frac{g}{(\gamma + 2k)^2 (\gamma + k)(1-S)} \left[ \Delta\tau - (1-S)\left(1 - e^{-\Delta\tau}\right) \right], \tag{40}$$

where the term in brackets is dimensionless. These definitions allow us to define the natural length unit:
$$L = \sqrt{\frac{g}{(\gamma + 2k)^2 (\gamma + k)(1-S)}}. \tag{41}$$

By rescaling the $MSD$, we define the dimensionless quantity,
$$\langle |\vec{\rho}|^2 \rangle = \frac{MSD}{\frac{g}{m^2(\gamma+2k)^2(\gamma+k)(1-S)}}, \tag{42}$$

and finally derive the mean-squared displacement in natural units:
$$\langle |\vec{\rho}|^2 \rangle = \Delta\tau - (1-S)\left(1 - e^{-\Delta\tau}\right), \tag{43}$$

which is exactly the modified Fürth Equation, proposed by Thomas and collaborators ( Ref. [9] in the main text). Observe that, if $k = 0$ in Eq. (40), we have:
$$MSD = \frac{g}{\gamma^2}\left[\Delta T - \frac{1}{\gamma}\left(1 - e^{-\gamma\Delta T}\right)\right], \tag{44}$$

that is, the original Fürth equation.

Note that, in the $\Delta T \to 0$ limit, Eq. (36) becomes:
$$MSD \approx 2qk\Delta T, \tag{45}$$

while for $\Delta T \to \infty$ limit, it becomes:
$$MSD \approx \frac{g(1+Q)}{(\gamma + 2k)(\gamma + k)}\Delta T. \tag{46}$$

## Velocity Auto-Correlation Functions

*VACF*

We can obtain the velocity auto-correlation function by calculating the product between two velocities for all time intervals and taking the average with respect to time of all the summed terms. Here we will consider only the velocity in the direction of the cell polarization; in this instantaneous direction it is well defined.

Assuming that $T + \Delta T = (n + \Delta n)\Delta t$:
$$\begin{aligned} VACF(\Delta T) &= \langle v_{||}(T+\Delta T)\vec{p}(T+\Delta T) \cdot v_{||}(T)\vec{p}(T)\rangle \\ &= \langle v_{||}((n+\Delta n)\Delta t)\vec{p}((n+\Delta n)\Delta t) \cdot v_{||}(n\Delta t)\vec{p}(n\Delta t)\rangle. \end{aligned} \tag{47}$$

Observe that here $\Delta T$ is not infinitesimal quantity like $\Delta t$. We will eventually take the limit $\Delta n \to \infty$, so that when $\Delta t \to 0$, $\Delta T$ remains finite.

Using Eq. (2) for $T$ and $\Delta T$, we obtain:

$$
\begin{aligned}
VACF(\Delta T) &= \Bigg\langle \Bigg[ v_{\|0}\Big(1-\gamma\Delta t\Big)^n [\vec{p}_0 \cdot \vec{p}(\Delta t)][\vec{p}(\Delta t) \cdot \vec{p}(2\Delta t)]\ldots[\vec{p}((n-1)\Delta t) \cdot \vec{p}(n\Delta t)]\vec{p}(n\Delta t) \\
&\quad + \int_0^{\Delta t} ds\, \xi_\| \Big(1-\gamma\Delta t\Big)^{n-1} [\vec{p}_0 \cdot \vec{p}(\Delta t)]\ldots[\vec{p}((n-1)\Delta t) \cdot \vec{p}(n\Delta t)]\vec{p}(n\Delta t) \\
&\quad + \int_{\Delta t}^{2\Delta t} ds\, \xi_\| \Big(1-\gamma\Delta t\Big)^{n-2} [\vec{p}(\Delta t) \cdot \vec{p}(2\Delta t)]\ldots[\vec{p}((n-1)\Delta t) \cdot \vec{p}(n\Delta t)]\vec{p}(n\Delta t) \\
&\quad + \ldots \\
&\quad + \int_{(n-1)\Delta t}^{n\Delta t} ds\, \xi_\| [\vec{p}((n-1)\Delta t) \cdot \vec{p}(n\Delta t)]\vec{p}(n\Delta t) \Bigg] \\
&\quad \cdot \Bigg[ v_{\|0}\Big(1-\gamma\Delta t\Big)^{n+\Delta n} [\vec{p}_0 \cdot \vec{p}(\Delta t)]\ldots[\vec{p}((n-1)\Delta t) \cdot \vec{p}(n\Delta t)][\vec{p}(n\Delta t) \cdot \vec{p}((n+1)\Delta t)] \\
&\quad \ldots [\vec{p}((n+\Delta n - 1)\Delta t) \cdot \vec{p}((n+\Delta n)\Delta t)]\vec{p}((n+\Delta n)\Delta t) \\
&\quad + \int_0^{\Delta t} ds\, \xi_\|(s) \Big(1-\gamma\Delta t\Big)^{n-1+\Delta n} [\vec{p}_0 \cdot \vec{p}(\Delta t)]\ldots[\vec{p}((n-1)\Delta t) \cdot \vec{p}(n\Delta t)][\vec{p}(n\Delta t) \cdot \vec{p}((n+1)\Delta t)] \\
&\quad \ldots [\vec{p}((n+\Delta n - 1)\Delta t) \cdot \vec{p}((n+\Delta n)\Delta t)]\vec{p}((n+\Delta n)\Delta t) \\
&\quad + \ldots \\
&\quad + \int_{(n-1)\Delta t}^{n\Delta t} ds\, \xi_\|(s) \Big(1-\gamma\Delta t\Big)^{\Delta n} [\vec{p}_0 \cdot \vec{p}(\Delta t)]\ldots[\vec{p}((n-1)\Delta t) \cdot \vec{p}(n\Delta t)][\vec{p}(n\Delta t) \cdot \vec{p}((n+1)\Delta t)] \\
&\quad \ldots [\vec{p}((n+\Delta n - 1)\Delta t) \cdot \vec{p}((n+\Delta n)\Delta t)]\vec{p}((n+\Delta n)\Delta t) \\
&\quad + \ldots \\
&\quad + \int_{(n-1)\Delta t}^{n\Delta t} ds\, \xi_\|(s) \Big(1-\gamma\Delta t\Big)^{\Delta n} [\vec{p}((n+\Delta n - 1)\Delta t) \cdot \vec{p}((n+\Delta n)\Delta t)]\vec{p}((n+\Delta n)\Delta t) \Bigg] \Bigg\rangle .
\end{aligned}
\quad (48)
$$

From the term products and averaging the integral terms, we find:

$$
\begin{aligned}
VACF(\Delta T) &= v_{\|0}^2 \Big(1-\gamma\Delta t\Big)^{2n+\Delta n} \Big\langle [\vec{p}_0 \cdot \vec{p}(\Delta t)]^2 \ldots [\vec{p}((n-1)\Delta t) \cdot \vec{p}(n\Delta t)]^2 \\
&\quad \cdot [\vec{p}(n\Delta t) \cdot \vec{p}((n+1)\Delta t)] \ldots [\vec{p}(n\Delta t) \cdot \vec{p}((n+\Delta n)\Delta t)] \Big\rangle \\
&\quad + g \int_0^{\Delta t} ds\, \xi_\|(s) \Big(1-\gamma\Delta t\Big)^{2(n-1)+\Delta n} \Big\langle [\vec{p}_0 \cdot \vec{p}(\Delta t)]^2 \ldots [\vec{p}((n-1)\Delta t) \cdot \vec{p}(n\Delta t)]^2 \\
&\quad \cdot [\vec{p}(n\Delta t) \cdot \vec{p}((n+1)\Delta t)] \\
&\quad \ldots [\vec{p}(n\Delta t) \cdot \vec{p}((n+\Delta n)\Delta t)] \Big\rangle \\
&\quad + \ldots \\
&\quad + g \int_{(n-1)\Delta t}^{n\Delta t} ds\, \xi_\|(s) \Big(1-\gamma\Delta t\Big)^{\Delta n} \Big\langle [\vec{p}(n\Delta t) \cdot \vec{p}((n+\Delta n)\Delta t)] \Big\rangle .
\end{aligned}
\quad (49)
$$

The cross terms resulting from the multiplication vanish because the product of two white noise terms at different instants has zero correlation. Applying Eq. (6), that is,

$$\langle \cos \Delta \theta_{i,j} \rangle_{\beta_\perp} \approx (1-k\Delta t)^{|j-i|} ,$$

and making the same assumptions we used to obtain Eq. (14), we obtain the equation:

$$
\begin{aligned}
VACF(\Delta T) &= v_{\|0}^2 \Big(1-\gamma\Delta t\Big)^{2n+\Delta n} (1-k\Delta t)^{2(n-1)+2\Delta n} \\
&\quad + g\Bigg[ \Big(1-\gamma\Delta t\Big)^{2(n-1)+\Delta n} (1-k\Delta t)^{2(n-1)+2\Delta n} + \ldots \\
&\quad + \Big(1-\gamma\Delta t\Big)^{\Delta n} (1-k\Delta t)^{2\Delta n - 1/2} \Bigg].
\end{aligned}
\quad (50)
$$

Since $v_{\|0}^2 = \frac{g}{2(\gamma+k)}$, we can conclude that:

$$VACF(\Delta T) = \Big(1-\gamma\Delta t\Big)^{\Delta n} (1-k\Delta t)^{2\Delta n - 1/2} \langle v_\|^2(T) \rangle . \quad (51)$$

Since $T = n\,\Delta t$, $\Delta T = \Delta n\,\Delta t$ and $\Delta n \to \infty$ we find:

$$\begin{aligned} VACF(\Delta T) &= \langle v_{||}^2(T)\rangle \left(1 - \gamma\frac{\Delta T}{\Delta n}\right)^{\Delta n}\left(1 - k\frac{\Delta T}{\Delta n}\right)^{2\Delta n - 1/2} \\ &= \langle v_{||}^2(T)\rangle\, e^{-(\gamma + 2k)\Delta T}\,. \end{aligned} \qquad (52)$$

**Mean Velocity Auto-Correlation Function $\psi^*(\Delta T, \varepsilon)$**

We defined the mean velocity as:
$$\overline{\vec{v}(T,\varepsilon)} = \frac{\vec{r}(T+\varepsilon) - \vec{r}(T)}{\varepsilon}, \qquad (53)$$

Observe that Eq. (53) includes both parallel and perpendicular displacements. We define the average velocity auto-correlation function as:
$$\psi^*(\Delta T, \varepsilon) = \left\langle \overline{\vec{v}(T,\varepsilon)} \cdot \overline{\vec{v}(T+\Delta T,\varepsilon)} \right\rangle. \qquad (54)$$

To evaluate Eq. (54) we must partition the time intervals involved. We write all time intervals as multiples of the infinitesimal interval $\Delta t$, and use the following convention:

$$\begin{aligned} T + \varepsilon &= T + n\Delta t, \\ T + \Delta T &= T + (m+n)\Delta T, \\ T + \Delta T + \varepsilon &= T + (m+2n)\Delta T\,. \end{aligned} \qquad (55)$$

Adapting Eq. (20) to displacements in the intervals $[T, T+\varepsilon]$ and $[T+\Delta T, T+\Delta T+\varepsilon]$, we verify that all cross terms have either $\xi_{||}$ or $\xi_\perp$ individually, or a product of the two. All these products vanish due to the average over noise implicit in the definition of $\psi^*(\Delta T, \varepsilon)$. After averaging over noise, we have:

$$\begin{aligned} \psi^*(\Delta T, \varepsilon) &= \frac{\Delta t^2}{\varepsilon^2}\langle v_{||}(T+\Delta T)v_{||}(T)\rangle \Big\langle \sum_{j=1}^{n}(1-\gamma\Delta t)^{n-j}(1-k\Delta t)^{n-j}\vec{p}_{m+2n-j} \\ &\quad \cdot \sum_{i=1}^{n}(1-\gamma\Delta t)^{n-i}(1-k\Delta t)^{n-i}\vec{p}_{n-i}\Big\rangle \\ &= \frac{\Delta t^2}{\delta^2}\langle v_{||}(T+\Delta T)v_{||}(T)\rangle\Big\langle \sum_{j=1}^{n}\sum_{i=1}^{n}(1-\gamma\Delta t)^{2n-j-i}(1-k\Delta t)^{2n-j-i}\vec{p}_{m+2n-j}\cdot\vec{p}_{n-i}\Big\rangle. \end{aligned} \qquad (56)$$

Since $\langle \vec{p}_0 \cdot \vec{p}_{m+n}\rangle = (1-k\Delta t)^{m+n}$ we write:

$$\psi^*(\Delta T, \varepsilon) = \frac{\Delta t^2}{\delta^2}\langle v_{||}(T+\Delta T)\vec{p}_{m+n}\cdot v_{||}(T)\vec{p}_0\rangle \sum_{j=1}^{n}(1-\gamma\Delta t)^{n-j}(1-k\Delta t)^{2(n-j)}\,. \qquad (57)$$

Now, taking $n \to \infty$ keeping $\varepsilon$ finite, and noting that the sums in the above equations are sums of geometrical series,

$$\psi^*(\Delta T, \varepsilon) = \frac{(1-e^{-\gamma\varepsilon})(1-e^{-(\gamma+2k)\varepsilon})}{\varepsilon^2\gamma(\gamma+2k)} VACF(\Delta T)\,. \qquad (58)$$

Observe that:
$$\lim_{\varepsilon \to 0} \psi(\Delta T, \delta) = VACF(\Delta T)\,, \qquad (59)$$

as it should. From Eq. (59), we can straightforwardly obtain the mean velocity auto-correlation function $\psi(\Delta\tau, \delta)$ given in natural units using Eqs. (39) and (41).